\documentclass[a4paper,11pt,prd,preprint, amsfonts,amsmath,amssymb,nofootinbib]{revtex4}
\usepackage[margin=15truemm]{geometry}
\usepackage{bm}
\usepackage[dvipdfmx]{graphicx}
\usepackage{ascmac}
\usepackage{amsthm}
\usepackage{latexsym}
\usepackage{physics}
\usepackage{color,float}
\usepackage{hyperref}
\usepackage{mathtools}
\usepackage{simplewick}
\usepackage{tikz}
\usepackage{adjustbox}
\usepackage{fancybox}
\usepackage{caption}
\usepackage{subcaption} 
\usepackage{tikz-feynman}
\usepackage{enumitem}
\usepackage{titlesec}
\begin{document}

\title{Effective description of open quantum dynamics in relativistic scattering}

\author{Kaito Kashiwagi} 
\email{kashiwagi.kaito.268@s.kyushu-u.ac.jp}
\author{Akira Matsumura}
\email{matsumura.akira@phys.kyushu-u.ac.jp}
\affiliation{Department of Physics, Kyushu University, Fukuoka, 819-0395, Japan}

\vspace{1cm}

\begin{abstract}
The open dynamics of quantum particles in relativistic scattering is investigated.
In particular, we consider the scattering process of quantum particles coupled to an environment initially in a vacuum state. 
Tracing out the environment and using the unitarity of S-operator, we find the Gorini-Kossakowski-Sudarshan-Lindblad (GKSL) generator describing the evolution of the particles.
The GKSL generator is exemplified by focusing on the concrete
processes: one is the decay of scalar particle ($\phi \rightarrow \chi \chi$), and the others are the pair annihilation and the $2\rightarrow 2$ scattering of scalar particles ($\phi \phi \rightarrow \chi \chi$ and $\phi \phi \rightarrow \phi \phi$). 
The GKSL generator for $\phi \rightarrow \chi \chi$ has a parameter with  the coupling between $\phi$ and $\chi$ and the mass of both fields. 
The GKSL generator associated with $\phi \phi \rightarrow \chi \chi$ is characterized by a Lorentz-invariant function of initial momenta. 
Especially, in the pair annihilation process, we show that the probability of pair annihilation varies depending on the superposition state of incident scalar $\phi$ particles.
Furthermore, we observe that the GKSL generators derived in this paper have Poincar\'e symmetry. This means that 
the description by 
the GKSL generator with Poincar\'e symmetry is effective for the asymptotic behavior of open quantum dynamics in the long-term processes of interest. 
\end{abstract}
\maketitle
\tableofcontents
\section{Introduction}
\label{sec:Intro}
The dynamics of quantum system described by a Shr\"odinger equation or a von Neumann equation assumes that the system is isolated.
However, isolated quantum systems do not strictly exist in reality.
It is because that the quantum system of interest interacts with an environment. 
In this case, the quantum system  
is called the open quantum system (OQS), and the theory of OQS has been developed\cite{Breuer2002}.
In the theory of OQS, a Gorini-Kossakowski-Sudarshan-Lindblad (GKSL) equation\cite{GKS_1976,Lindblad_1976} is effective for describing the Markovian dynamics of OQS and the dissipation of OQS.
For example, the equation can explain the spontaneous emission of two-level atom, the relaxation of spin and so on\cite{Breuer2002,Kimura_2008}.
Two-level atom and spin system are treated as non-relativistic systems, and the theory of OQS is well-established in the non-relativistic regime. 
However, we also meet the dissipative phenomena in the relativistic regime.
The pion decay is taken as the example.
For the decay of neutral pion into two photons, if we focus on the neutral pion system, its dynamics is dissipative because the pion state changes to a vacuum state.
This suggests that approaches of the theory of OQS may be effective in describing relativistic dissipative phenomena\cite{Alicki1986, Wang2022, Baidya2017, Meng2021, Minami_2018,Hongo_2021,Diosi_2022,Matsumura2023}.

It is natural to ask whether the theory of OQS is needed in explaining relativistic dissipative phenomena. 
We consider that there are two advantages to using the theory of OQS.
One is that the effective parameters for characterizing the dynamics of OQS can be obtained.
In the case of spontaneous emission, the decay rate of two-level atom is an example of such parameters \cite{Breuer2002}. 
It gives the timescale of dissipation due to emission without the detailed information of environment.
In this viewpoint, the theory of OQS can be applied to identify the effective theory of relativistic dissipative dynamics.

The other advantage is that the approaches of quantum information theory can be applied.
Since 1970s, the researches about Unruh effect\cite{Unruh_1976}, Hawking radiation\cite{Hawking_1975}, and related topics are actively pursued in the fusion area of general relativity and quantum information theory. 
For example, the spacetime symmetry restricts the invariance of quantum entanglement\cite{Shi_2004}, and the entanglement between two Unruh-Dewitt detectors harvesting is inhibited by black hole\cite{Henderson_2018}, and so on. These do not deal with OQS, but they imply that quantum entanglement is effective in characterizing relativistic phenomena in paradigm of the theory of OQS. We consider that a black hole is one of the good research subjects for the theory of OQS.
One of the challenges in this field is to resolve the black hole information loss paradox. For example, the process of information loss was discussed through the models of decoherence and recoherence in open quantum systems\cite{ALZP_1995}. 
Formulating the theory of relativistic OQS potentially gives a clue for resolving the information loss paradox.

Our work is also motivated by the challenge of constructing quantum gravity theory.
Quantum gravity theory has not been established and the experimental clues of the theory are not obtained.
This situation stimulates two theoretical standpoints: 
one is that ``gravity is quantum" and the other is that ``gravity is classical".
Some theories and models  are proposed in each standpoint. 
In the former standpoint, perturbative quantum gravity\cite{Donoghue_1994} was proposed as one of the typical theories. 
This theory adopts the canonical quantization of matter field and gravitational field in the weak regime. 
The evolution of fields predicted from the theory is unitary in the regime.
On the other hand, in the latter standpoint, the Di\'osi-Penrose model \cite{Diosi_1987,Diosi_1989,Penrose_1996} and the Kafri-Taylor-Milburn model \cite{KTM_2014} were proposed.
They predict the collapse of matter wavefunction due to gravity. 
The wavefunction collapse comes from the assumption that gravitational field does not obey the quantum superposition principle and is in a classically definite state. 
The non-relativistic dynamics of quantum matter coupled to the gravitational field is described by a GKSL equation. 
In addition, the relativistic theory of classical gravity coupled to quantum matter was also proposed\cite{Oppenheim_2023}.

If the above models and theories are unified within a theoretical paradigm, we can achieve a comprehensive approach for exploring them.
To establish the paradigm, we have been developing the theory of the GKSL equation with Poincaré symmetry\cite{Kaito_2024}.
We expect that the present study would contribute to the development of the paradigm. 
Notably, experiments aimed at verifying the models and theories have already been proposed, and rapid advancements in quantum technologies would realize the proposals \cite{Carney_2019}.
The theoretical paradigm unifying advocated models and theories would accelerate progress in quantum gravity research by combining future experimental efforts.
This serves as a key motivation for undertaking this study.

As we mentioned above, there are many advantages to describe the relativistic phenomena by the theory of OQS. In this paper, we consider the dynamics of two scalar fields $\phi$ and $\chi$ in a long timescale and show that the reduced dynamics of $\phi$ field can be described by GKSL generator. 
To show the explicit form of GKSL generator, we take the following processes: the decay of scalar particle ($\phi \rightarrow \chi \chi$),  the pair annihilation and the $2\rightarrow2$ scattering of scalar particles ($\phi\phi \rightarrow \chi\chi$ and $\phi \phi \rightarrow \phi \phi$). We then derive the GKSL generators associated with the processes. 
In the process of the decay of scalar particle ($\phi \rightarrow \chi\chi$), we show that the GKSL generator has a decay parameter which consists of the coupling constant and the masses of both fields, and we also find that this generator asymptotically has Poincar\'e symmetry and describes inevitable dissipation. 
In the processes of pair annihilation and scattering of scalar particles ($\phi\phi \rightarrow \chi\chi$ and $\phi\phi \rightarrow \phi\phi$), it turns out that the generator also has Poincar\'e symmetry and describes the pair annihilation of scalar particles $\phi$ and the interaction by the exchange of virtual particles $\chi$.
In particular, we observe that the pair annihilation of scalar particles is characterized by a Lorentz-invariant function of initial momenta. 
We further see that the pair annihilation probability depends on the relative phase in the quantum superposition state of the incident particles.

The structure of this paper is as follow. In Sec.\ref{sec:formulation_GKSL}, we derive the abstract expression of GKSL generator, exemplify the GKSL generator for each scattering process and observe the properties of GKSL generator in Sec.\ref{sec:Examples_GKSL}. In Sec.\ref{sec:Discussion}, we discuss the properties of GKSL generator in this study and the relation with previous research. Finally, we conclude this study and mention the future outlook in Sec.\ref{sec:conclusion and outlook}. 
In the following we adopt the natural unit $\hbar=c=1$ and the convention of the Minkowski metric as
$\eta_{\mu \nu}=\text{diag}[-1,1,1,1]$.
The commutator and the anticommutator are defined as 
$[\hat{A},\hat{B}]=\hat{A}\hat{B}-\hat{B}\hat{A}$ and
$\{\hat{A},\hat{B}\}=\hat{A}\hat{B}+\hat{B}\hat{A}$, respectively.
\section{Formulation by the theory of Open Quantum System}
\label{sec:formulation_GKSL}
The theory of OQS describes the dissipative dynamics of quantum system interacting with an environment. 
As mentioned in Introduction, the dynamics can be often governed by a GKSL equation.
As is written in Ref.\cite{Breuer2002}, the GKSL equation for the reduced density operator $\rho_{s}(t)$ of an OQS is derived by the three approximations: the Born approximation, the Markov approximation, and the rotating wave approximation. 
The GKSL equation has the form,
\begin{align}
    \frac{d}{dt} \rho_{\mathrm{s}}(t) = \mathcal{L} \left[ \rho_{\mathrm{s}}(t) \right], 
    \label{eq:rho_s}
\end{align}
where the GKSL generator $\mathcal{L}$ is 
\begin{align}
    \mathcal{L}\left[ \rho_{\mathrm{s}}(t) \right] 
    = -i\left[ \hat{M}, \rho_{\mathrm{s}}(t) \right] +  \sum _{\lambda} \left[ \hat{L}_{\lambda} \rho_{\mathrm{s}}(t) \hat{L}^{\dagger}_{\lambda} - \frac{1}{2} \left\{ \hat{L}^{\dagger}_{\lambda} \hat{L}_{\lambda} , \rho_{\mathrm{s}}(t) \right\} \right],
    \label{eq:L}
\end{align}
and the operators $\hat{M}$, $\hat{L}_\lambda$ are an Hermitian operator and a Lindblad operator, respectively. The Lindblad operator contains the information of environment and gives the non-unitary evolution of OQS.
In this section, we will derive a GKSL generator based on scattering theory.

For concreteness, we consider the model of two scalar fields $\phi$ and $\chi$ with the total Hamiltonian $\hat{H}_{ \mathrm{tot}}$,
\begin{align}
    \hat{H}_{ \mathrm{tot} } &=
    \hat{H}_{ \mathrm{s} } \otimes \hat{ \mathbb{I} }_{ \mathrm{E} } +
    \hat{\mathbb{I}}_{ \mathrm{s} } \otimes \hat{H}_{ \mathrm{E} }
    + \hat{V},
    \label{eq:H_tot}
    \\
    \hat{H}_{ \mathrm{s} } &=
    \frac{1}{2} \int d^3 x \ \Big[ 
    \hat{\pi}^2_{ \phi } + (\nabla \hat{\phi})^2 + m_\text{s}^2 \hat{\phi}^2
    \Big],
    \label{eq:H_s}
    \\
    \hat{H}_{ \mathrm{E} } &=
    \frac{1}{2} \int d^3 x \ \Big[ 
    \hat{\pi}^2_{ \chi } + (\nabla \hat{\chi})^2 + m_\text{E}^2 \hat{\chi}^2
    \Big],
    \label{eq:H_E}
    \\
    \hat{V} &= - \lambda \int d^3 x \ \hat{\phi}(\bm{x}) \otimes \hat{\chi}^2 (\bm{x}), 
    \label{eq:exV}
\end{align}
where $\hat{\pi}_{ \phi }$, $\hat{\pi}_{ \chi }$ represent the conjugate momenta for each scalar field, and $\lambda$ is a coupling constant. 
In the following, we regard the scalar field $\phi$ as an OQS and the scalar field $\chi$ as an environment coupled to the OQS.
The subscript $\mathrm{s}$ labels the OQS and 
the subscript $\mathrm{E}$ labels the environment.
In scattering theory, an S-operator $\hat{S}$ gives the time evolution from an in-state $\rho^\text{in}_\text{tot}$ to an out-state $\rho^\text{out}_\text{tot}=\hat{S} \rho^\text{in}_\text{tot} \hat{S}^\dagger $. The S-operator is defined by using the interaction $\hat{V}_{\mathrm{I}}(t)$ in the interaction picture, 
\begin{equation}
    \hat{S} = \mathrm{T} \exp\left( -i \int^{\infty}_{-\infty} dt \ \hat{V}_{\mathrm{I}}(t) \right).
    \label{eq:S}
\end{equation} 
Here, taking the in-state 
$\rho^\text{in}_\text{tot}=\rho^\text{in}_\text{s} \otimes |0\rangle_\text{E} \langle 0|$, we consider the scattering dynamics:
\begin{align}
    \rho^\text{out}_{\mathrm{tot}}=  \hat{S} \rho^\text{in}_{\mathrm{s}} \otimes |0 \rangle _{\mathrm{E}} \langle 0| \hat{S}^{\dagger},
\end{align}
where $\rho^\text{in}_\text{s}$ is the initial density operator of the OQS and $|0\rangle_\text{E}$ is the vacuum state of environment. 
The initial condition is relevant for the decay process $(\phi \rightarrow \chi\chi)$ and the pair annihilation process $(\phi \phi \rightarrow \chi \chi)$. 
Tracing out the environment from the out state $\rho^\text{out}_\mathrm{tot}$, we have the reduced density operator of OQS as 
\begin{align}
 \rho^\text{out}_\mathrm{s} = \mathrm{Tr}_\text{E} \left[ \hat{S} \rho^\text{in}_{\mathrm{s}} \otimes |0 \rangle _{\mathrm{E}} \langle 0| \hat{S}^{\dagger} \right] 
    \label{eq:S-dynamics}
\end{align}
For convenience, we divide the S-operator into two parts, 
\begin{align}
    \hat{S} = \hat{\mathbb{I}} + i \hat{T},
    \label{eq:T-operator}
\end{align}
where $\hat{\mathbb{I}}$ reflects free evolution and $\hat{T}$ has all information of scattering processes.
Plugging this expression into Eq.\eqref{eq:S-dynamics}, we can rewrite it as follow:
\begin{align}
    \rho^\text{out}_{ \mathrm{s} }
    &= \rho^\text{in}_{ \mathrm{s} } + i \left[ \mathrm{Re} \hat{T}_{0}, \rho^\text{in}_\text{s}\right] + \int d\beta \ \left[ \hat{T}_{\beta} \rho^\text{in}_{ \mathrm{s} }\hat{T}^{\dagger}_{\beta} - \frac{1}{2} \left\{ \hat{T}^{\dagger}_{\beta} \hat{T}_{\beta}, \rho^\text{in}_{ \mathrm{s} } \right\}\right],
    \label{eq:S-dynamics2}
\end{align}
where $\text{Re}\hat{T}_0=(\hat{T}_0+\hat{T}^\dagger_0)/2$, and we defined the operators of OQS,
\begin{align}
    \hat{T}_{\beta} = {}_\text{E} \langle \beta | \hat{T} |0 \rangle_\text{E}, \quad \hat{T}_{0} = {}_\text{E} \langle 0| \hat{T} | 0 \rangle_\text{E}.
    \label{eq:def_T}
\end{align}
Here, $|\beta \rangle_\text{E}$ is the Fock state of environment, which satisfies the completeness condition on the Hibert space of environment, $\int d\beta |\beta\rangle_\text{E} \langle \beta|=|0\rangle_\text{E} \langle 0|+\int d^3\bm{p} |\bm{p}\rangle_\text{E} \langle \bm{p}|+\cdots=\hat{\mathbb{I}}_\text{E}$.
The detailed derivation of Eq.\eqref{eq:S-dynamics2} is presented in Appendix \ref{app:drv1}. 
In this derivation, we used the unitarity condition of the S-operator and the equation derived from it,
\begin{align}
    \hat{S}^{\dagger} \hat{S} = \hat{ \mathbb{I} } 
    \qquad \Longrightarrow
    \qquad 
    \mathrm{Im}\hat{T}_{0} = \frac{1}{2} \int d\beta \ \hat{T}^{\dagger}_{\beta} \hat{T}_{\beta},
    \label{eq:unitary_condition}
\end{align}
where $\text{Im}\hat{T}_0=(\hat{T}_0-\hat{T}^\dagger_0)/2i$. 
As can be seen, the second and third terms of \eqref{eq:S-dynamics2} is expressed by a GKSL generator. That is, denoting the change of state as $\rho^\text{out}_\text{s}-\rho^\text{in}_\text{s}=
\mathcal{L}_{ \mathrm{scatt} } \left[ \rho^\text{in}_{ \mathrm{s} } \right]$, we find the following generator in the scattering theory:
\begin{align}
    \mathcal{L}_{ \mathrm{scatt} } \left[ \rho^\text{in}_{ \mathrm{s} } \right]
    = i \left[ \mathrm{Re} \hat{T}_{0}, \rho^\text{in}_{s} \right] + \int d\beta \ \left[ \hat{T}_{\beta} \rho^\text{in}_{s} \hat{T}^{\dagger}_{\beta} - \frac{1}{2} \left\{ \hat{T}^{\dagger}_{\beta} \hat{T}_{\beta}, \rho^\text{in}_{s} \right\}\right],
    \label{eq:L_scatt}
\end{align}
where $\text{Re}\hat{T}_0$ and $\hat{T}_\beta$ defined through Eq.\eqref{eq:def_T} give the unitary evolution and the non-unitary evolution of OQS, respectively.  
According to the definition \eqref{eq:def_T}, the Hermitian operator $\text{Re}\hat{T}_0$ reflects the process during which the environment remains in the vacuum state. The Lindblad operator $\hat{T}_\beta$ potentially comes from the excitation of the environment.

What we want to emphasize here is that one does not need the explicit form of S-operator to get the generator. 
The derivation relies only on two assumptions: the initial state of environment is vacuum and the S-operator is unitary.
In this sense, the above GKSL generator generally emerges for describing open quantum dynamics in scattering processes. 
The operators $\mathrm{Re}\hat{T}_{0}$ and $\hat{T}_{\beta}$ are calculated from Feynman diagrams, and we can obtain the expression of the GKSL generator $\mathcal{L}_{\mathrm{scatt}}$ considering concrete scattering processes. The physical reason why we can get the GKSL generator will be presented in Sec.\ref{sec:Discussion} after demonstrating the specific expression of Eq.\eqref{eq:L_scatt}. 


\section{Examples of GKSL generator}
\label{sec:Examples_GKSL}
In this section, we consider the following processes: the decay of scalar particle $(\phi\rightarrow \chi \chi)$, the pair annihilation of scalar particles $(\phi\phi \rightarrow \chi \chi)$ and the $2\rightarrow 2$ scattering of scalar particles ($\phi \phi \rightarrow \phi \phi$). 
We then derive the examples of GKSL generator for the processes. 
Note that we regard the scalar field $\phi$ as an OQS and the scalar field $\chi$ as an environment, and the interaction between them is given as Eq.\eqref{eq:exV}. 
Hereafter, we use the following Feynman rules based on Ref.\cite{Weinberg_1995}.
\begin{itembox}[c]{Feynman rules in momentum space}
    \begin{enumerate}
        \item For each vertex, the factor $-i\lambda (2\pi)^4$ and delta function reflecting the four-momentum conservation law are assigned:
        \begin{equation*}
            (-i\lambda)(2\pi)^4 \delta^4(\sum p + \sum q - \sum p' - \sum q')
        \end{equation*}
        \item For each external dotted line, the factor $(2\pi)^{ -\frac{3}{2} } (2\omega_{\bm{p}})^{-\frac{1}{2}}$ shown in (a) is assigned, where $\omega_{\bm{p}}$ is the energy of particle $\phi$. For each external line, the factor $(2\pi)^{ -\frac{3}{2} } (2E_{\bm{k}})^{-\frac{1}{2}}$ shown in (a) is assigned, where $E_{\bm{k}}$ is the energy of particle $\chi$. 
        \item For each internal dotted line, the Feynman propagator $\dfrac{1}{(2\pi)^4} \dfrac{-i}{q^{2} + m_\text{s}^2 -i\epsilon}$ shown in (b) is assigned.
        For each internal line, the Feynman propagator $\dfrac{1}{(2\pi)^4} \dfrac{-i}{q^{2} + m_\text{E}^2 -i\epsilon}$ shown in (b) is assigned.
        \item Finally, we perform the integral with respect to all internal four-momentum.
    \end{enumerate}
\end{itembox}
\begin{center}
\begin{tabular}{cc} 
    \begin{minipage}{0.5\textwidth}
    \centering
    \begin{adjustbox}{valign=c}
    \begin{tikzpicture}[baseline=(current bounding box.center), scale=0.8]
        \begin{feynman}
            \vertex (a) at (0,2);
            \vertex (i) at (0,-1.5);
            \vertex (j) at (0.5,-2);
            \vertex (k) at (-0.5,-2);
            \vertex (o) at (0,0);
            \diagram*{
            (i) --  (j),
            (i) --  (k),
            (i) -- [scalar](o) -- [scalar,momentum=\(p\)] (a),
            };
            \fill (i) circle (2pt);
        \end{feynman}
    \end{tikzpicture}
    \end{adjustbox}
    $ = \dfrac{ (2\pi)^{ - \frac{3}{2} } }{\sqrt{2\omega_{\bm{p}}}} $ \\
    \vspace{1cm}
    \end{minipage}
    &
    \begin{minipage}{0.5\textwidth}
    \centering
    \begin{adjustbox}{valign=c}
    \begin{tikzpicture}[baseline=(current bounding box.center), scale=0.8]
        \begin{feynman}
            \vertex (a) at (0,2);
            \vertex (i) at (0,-1.5);
            \vertex (j) at (0.5,-2);
            \vertex (k) at (-0.5,-2);
            \vertex (i') at (0,1.5);
            \vertex (j') at (0.5,2);
            \vertex (k') at (-0.5,2);
            \vertex (o) at (0,0);
            \diagram*{
            (i) --  (j),
            (i) --  (k),
            (i) -- [scalar,momentum=\(q\)] (i'),
            (i') -- (j'),
            (i') -- (k'),
            };
            \fill (i) circle (2pt);
            \fill (i') circle (2pt);
        \end{feynman}
    \end{tikzpicture}
    \end{adjustbox}
    $ = \dfrac{1}{(2\pi)^4} \dfrac{-i}{q^{2} + m_\text{s}^2 -i\epsilon} $ \\
    \vspace{1cm}
    \end{minipage}
    \\
    \begin{minipage}{0.5\textwidth}
    \centering
    \begin{adjustbox}{valign=c}
    \begin{tikzpicture}[baseline=(current bounding box.center), scale=0.8]
        \begin{feynman}
            \vertex (a) at (0,2);
            \vertex (i) at (0,-1.5);
            \vertex (j) at (0.5,-2);
            \vertex (k) at (-0.5,-2);
            \vertex (o) at (0,0);
            \diagram*{
            (i) --  (j),
            (i) -- [scalar](k) ,
            (i) -- (o) -- [momentum=\(k\)] (a),
            };
            \fill (i) circle (2pt);
        \end{feynman}
    \end{tikzpicture}
    \end{adjustbox}
    $ = \dfrac{ (2\pi)^{ - \frac{3}{2} } }{\sqrt{2E_{\bm{k}}}} $ \\
    \vspace{1cm}
    \text{(a) : External line}
    \end{minipage}
    &
    \begin{minipage}{0.5\textwidth}
    \centering
    \begin{adjustbox}{valign=c}
    \begin{tikzpicture}[baseline=(current bounding box.center), scale=0.8]
        \begin{feynman}
            \vertex (a) at (0,2);
            \vertex (i) at (0,-1.5);
            \vertex (j) at (0.5,-2);
            \vertex (k) at (-0.5,-2);
            \vertex (i') at (0,1.5);
            \vertex (j') at (0.5,2);
            \vertex (k') at (-0.5,2);
            \vertex (o) at (0,0);
            \diagram*{
            (i) --  (j),
            (i) --  [scalar](k),
            (i) -- [momentum=\(q\)] (i'),
            (i') -- (j'),
            (i') -- [scalar] (k'),
            };
            \fill (i) circle (2pt);
            \fill (i') circle (2pt);
        \end{feynman}
    \end{tikzpicture}
    \end{adjustbox}
    $ = \dfrac{1}{(2\pi)^4} \dfrac{-i}{q^{2} + m_\text{E}^2 -i\epsilon} $ \\
    \vspace{1cm}
    \text{(b) : Internal line}
    \end{minipage}
\end{tabular}
\label{tab:Feynman_rule}
\end{center}


\subsection{Decaying scalar particle}
\label{sec:Decay_process}

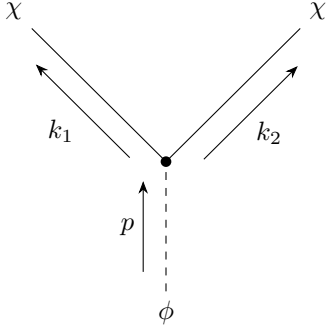
\begin{figure}[H]
    \centering
    \begin{tikzpicture}
    \definecolor{darkgreen}{RGB}{0,100,0}
    \begin{feynman}
      \vertex (a) at (0,-2) {$\phi$};
      \vertex (b) at (-2,2) {$\chi$};
      \vertex (c) at (2,2) {$\chi$};
      \vertex (i) at (0,0);
      \diagram* {
        (a) -- [scalar, momentum=\(p\)] (i),
        (i) -- [momentum=\(k_1\)] (b),
        (i) -- [momentum={[arrow distance=-3.3mm, label distance=-15pt] \(k_2\)}] (c),
      };
      \fill (i) circle (2pt);
    \end{feynman}
  \end{tikzpicture}
  \caption{Decaying scalar particle}
  \label{fig:Decay_model}
\end{figure}

In this subsection, we will get the GKSL generator for the decay process of a single particle ($\phi \rightarrow \chi \chi$). 
We first compute the expression of the Lindblad operator $\hat{T}_{\beta}$. 
According to Fig.\ref{fig:Decay_model}, it is sufficient to expand Eq.\eqref{eq:S} up to the first order of $\lambda$. 
Up to this order, the S-operator is given as
\begin{align}
    \hat{S} \approx \hat{ \mathbb{I} } + i \lambda \int d^4 x \ \hat{\phi}(t,\bm{x}) \otimes \hat{\chi}^2(t,\bm{x}), 
    \label{eq:S_decay}
\end{align}
and then the T-operator is
\begin{align}
    \hat{T} = \lambda \int d^4 x \ \hat{\phi}(t,\bm{x}) \otimes \hat{\chi}^2(t,\bm{x}).
    \label{eq:T_decay}
\end{align}
Using the completeness condition on the Hilbert space of OQS, $\int d\alpha |\alpha\rangle_\text{s} \langle \alpha|= |0\rangle_\text{s} \langle 0| + \int d^3p |\bm{p} \rangle_\text{s}\langle \bm{p}| + \cdots = \hat{\mathbb{I}}_\text{s}$, we can rewrite $\hat{T}_{\beta}$ as follows:
\begin{align}
    \hat{T}_{\beta} 
    = \int d \bar{\alpha} d \alpha \left[ {}_{\mathrm{E}}\langle \beta | {}_{\mathrm{s}}\langle \bar{\alpha} | \ \hat{T} \ | \alpha \rangle_{\mathrm{s}}  |0 \rangle_{\mathrm{E}} \right] | \bar{\alpha} \rangle _{\mathrm{s}} \langle \alpha |.
    \label{eq:T^s_beta1}
\end{align}
In proceeding with this calculation, the derivation of the scattering amplitude by the Feynman rules is useful.
The amplitude is computed from the Feynman diagram Fig.\ref{fig:Decay_model} with one vertex and three external lines. 
In this case, we have the  scattering amplitude ${}_{\mathrm{E}}\langle \beta |  {}_{\mathrm{s}}\langle \bar{\alpha} | \ i\hat{T} \ | \alpha \rangle_{\mathrm{s}}  |0 \rangle_{\mathrm{E}} $ reflecting the transition from the state $ | \alpha \rangle_\text{s}  |0 \rangle_\text{E} = | \bm{p} \rangle_\text{s} | 0 \rangle_\text{E} $ to the state $ | \bar{ \alpha } \rangle_\text{s} | \beta \rangle_\text{E} = |0 \rangle_\text{s} | \bm{k}_1, \bm{k}_2 \rangle_\text{E} $.
It is explicitly given as
\begin{align}
   {}_{\mathrm{E}}\langle \beta | {}_{\mathrm{s}}\langle \bar{\alpha} | \ i\hat{T} \ | \alpha \rangle_{\mathrm{s}}  |0 \rangle_{\mathrm{E}}
    &= \frac{ (2\pi)^{ - \frac{3}{2} } }{ \sqrt{ 2E_{\bm{k}_1} } } 
    \cdot \frac{ (2\pi)^{ - \frac{3}{2} } }{ \sqrt{ 2E_{\bm{k}_2} } }
    \cdot \frac{ (2\pi)^{ - \frac{3}{2} } }{ \sqrt{ 2\omega_{\bm{p}} } }
    \cdot (-i \lambda) (2\pi)^4 \delta^4 (k_1 + k_2 -p) \notag \\
    &= - \frac{ i\lambda }{ 2\sqrt{ \pi E_{\bm{k}_1} E_{\bm{k}_2} \omega_{\bm{p}} } }
    \delta^4 (k_1 + k_2 -p),
    \label{eq:S_decay2}
\end{align}
where $E_{\bm{k}}=\sqrt{\bm{k}^2 + m_\text{E}^2}$ and $\omega_{\bm{p}} = \sqrt{ \bm{p}^2 + m_\text{s}^2 }$. Substituting this for Eq.\eqref{eq:T^s_beta1}, we obtain the Lindblad operator $\hat{T}_{\beta}=\hat{T}_{\bm{k}_1 \bm{k}_2}$,
\begin{align}
    \hat{T}_{\bm{k}_1 \bm{k}_2} 
    = -\frac{\lambda}{2\sqrt{\pi E_{\bm{k}_1} E_{\bm{k}_2}}} \int  \frac{d^3p}{\sqrt{\omega_{\bm{p}}}}\delta^4 (k_1 + k_2 -p)\ | 
    0 \rangle _{\mathrm{s}} \langle \bm{p} |,
    \label{eq:T^s_beta2}
\end{align}

Next, we derive the $\hat{T}_0$ to compute $\mathrm{Re} \hat{T}_0$. Following the same procedure as in the case of $\hat{T}_{\beta}$, the definition of $\hat{T}_0$ can be rewritten as
\begin{align}
    \hat{T}_0 
    &= \int d \bar{\alpha} d \alpha  \left[ {}_{\mathrm{E}}\langle 0 | {}_{\mathrm{s}}\langle \bar{\alpha} | \ \hat{T} \ | \alpha \rangle_{\mathrm{s}} |0 \rangle_{\mathrm{E}} \right] | \bar{\alpha} \rangle _{\mathrm{s}} \langle \alpha |,
    \label{eq:T^s_0-1}
\end{align}
In this case, the scattering amplitude is ${}_{\mathrm{E}}\langle 0 | {}_{\mathrm{s}}\langle \bar{\alpha} | \ i\hat{T} \ | \alpha \rangle_{\mathrm{s}} |0 \rangle_{\mathrm{E}}$. 
As can be seen in Eqs.\eqref{eq:L_scatt} and \eqref{eq:T^s_beta2}, it turns out that the second term in Eq.\eqref{eq:L_scatt} has the second order of $\lambda$. 
Namely, we must consider up to $O(\lambda^2)$ in calculating the first term with $\mathrm{Re} \hat{T}_0$ in Eq.\eqref{eq:L_scatt}. 
The amplitude is given by the Feynman diagram shown in Fig.\ref{fig:one-loop}, which describes the transition from $ | \alpha \rangle_\text{s}  |0 \rangle_\text{E} = | \bm{p} \rangle_\text{s} | 0 \rangle_\text{E} $ to $ | \bar{ \alpha } \rangle_\text{s} | 0 \rangle_\text{E} = |\bar{\bm{p}} \rangle_\text{s} |0 \rangle_\text{E} $. 
The diagram has an ultraviolet (UV)  divergence, but it can be removed by renormalizing the mass $m_\text{s}$ of the field $\phi$.  
We provide the renormalization procedure in Appendix \ref{app:renormalization}, and the value of ${}_{\mathrm{E}}\langle 0 | {}_{\mathrm{s}}\langle \bar{\bm{p}} | \ i\hat{T} \ | \bm{p} \rangle_{\mathrm{s}} |0 \rangle_{\mathrm{E}}$ is set to zero by the mass counterterm.
Of course, there exist other Feynman diagrams shown in Fig.\ref{fig:vacuum1}.
However, we do not have to consider those disconnected diagrams because they are vacuum bubble diagrams and can be absorbed by the redefinition of S-operator. 
The detailed discussion is presented in Appendix \ref{app:Re_Soperator}.

\begin{figure}[H]
  \begin{minipage}[b]{0.33\textwidth}
    \centering
    \begin{tikzpicture} 
    \begin{feynman}
      \vertex (a) at (0,-2) {$\phi$};
      \vertex (b) at (0,2) {$\phi$};
      \vertex (i) at (0,-1);
      \vertex (f) at (0,1);
      \vertex (n1) at (1,0) [label=right:{$\chi$}] ;
      \vertex (n2) at (-1,0) [label=left:{$\chi$}] ;

      \diagram* {
        (a) -- [scalar, momentum=\(p\)] (i),
        (i) -- [quarter right] (n1) -- [quarter right] (f) -- [quarter right] (n2) -- [quarter right] (i) ,
        (f) -- [scalar, momentum=\( \bar{p} \)] (b),
      };
      \fill (i) circle (2pt);
      \fill (f) circle (2pt);
    \end{feynman}
  \end{tikzpicture}
  \caption{One-loop diagram}
  \label{fig:one-loop}
  \end{minipage}
  \begin{minipage}[b]{0.66\textwidth}
  \centering
    \begin{minipage}[b]{0.33\textwidth}
    \centering
    \begin{tikzpicture}
    \definecolor{darkgreen}{RGB}{0,100,0}
    \begin{feynman}
      \vertex (a) at (0,-2) {$\phi$};
      \vertex (b) at (0,2) {$\phi$};
      \vertex (j1) at (0, -0.5) ;
      \vertex (j2) at (0, 0.5) ;
      \vertex (i1) at (1,0)  ;
      \vertex (i2) at (2,1) [label=above:{$\chi$}];
      \vertex (i3) at (3,0) ;
      \vertex (i4) at (2,-1) [label=below:{$\chi$}];

      \diagram* {
        (a) -- [scalar, momentum=\(p\)] (j1) -- [scalar] (j2) -- [scalar, momentum=\(\bar{p}\)] (b),
        (i1) -- [quarter left] (i2) -- [quarter left] (i3) -- [quarter left] (i4) -- [quarter left] (i1),
        (i1) -- [scalar, edge label'=\(\phi\)] (i3),
      };
      \fill (i1) circle (2pt);
      \fill (i3) circle (2pt);
    \end{feynman}
  \end{tikzpicture}
  \end{minipage}
  \hspace{1.5cm}
  \begin{minipage}[b]{0.33\textwidth}
    \centering
    \begin{tikzpicture}
    \definecolor{darkgreen}{RGB}{0,100,0}
    \begin{feynman}
      \vertex (a) at (0,-2) {$\phi$};
      \vertex (b) at (0,2) {$\phi$};
      \vertex (j1) at (0, -0.5) ;
      \vertex (j2) at (0, 0.5) ;
      \vertex (i1) at (0.5,0) ;
      \vertex (i2) at (1.0,0.5) [label=above:{$\chi$}];
      \vertex (i3) at (1.5,0) ;
      \vertex (i4) at (1.0,-0.5) [label=below:{$\chi$}];
      \vertex (k1) at (2.5,0) ;
      \vertex (k2) at (3.0,0.5) [label=above:{$\chi$}];
      \vertex (k3) at (3.5,0) ;
      \vertex (k4) at (3.0,-0.5) [label=below:{$\chi$}];

      \diagram* {
        (a) -- [scalar, momentum=\(p\)] (j1) -- [scalar] (j2) -- [scalar, momentum=\(\bar{p}\)] (b),
        (i1) -- [quarter left] (i2) -- [quarter left] (i3) -- [quarter left] (i4) -- [quarter left] (i1),
        (k1) -- [quarter left] (k2) -- [quarter left] (k3) -- [quarter left] (k4) -- [quarter left] (k1),
        (i3) -- [scalar, edge label'=\(\phi\)] (k1),
      };
      \fill (k1) circle (2pt);
      \fill (i3) circle (2pt);
    \end{feynman}
  \end{tikzpicture}
  \end{minipage}
  \caption{Vacuum bubble diagrams}
  \label{fig:vacuum1}
  \end{minipage}
\end{figure}
Let $\mathcal{L}_\phi$ denote the GKSL generator in this process. Using $\hat{T}_{\bm{k}_1 \bm{k}_2}$, we can obtain the GKSL generator as
\begin{align}
    &\mathcal{L}_{\phi}\left[ \rho^\text{in}_{\mathrm{s}} \right]
    = \int \frac{d^3 p \, d^3 \bar{p}}{\sqrt{ \omega_{\bm{p}}\,\omega_{\bar{\bm{p}}} }} 
    \gamma(p)\delta^4 (p-\bar{p})
    \left[ \hat{ \mathrm{a} }_{ \bm{p} } \rho^\text{in}_{\mathrm{s}}\hat{ \mathrm{a} }^{\dagger}_{\bar{\bm{p}}} - \frac{1}{2} \left\{ \hat{\mathrm{a}}^\dagger_{\bm{ \bar{p}}} \hat{\mathrm{a}}_{\bm{p}}, \rho^\text{in}_{\mathrm{s}} \right\} \right], 
    \label{eq:L_decay}
    \\
    &\gamma(p) 
    = \frac{\lambda^2}{4\pi} \int \frac{d^3 k_1 \, d^3 k_2}{E_{\bm{k}_1} E_{\bm{k}_2}}  \delta^4 (p - k_1 - k_2),
    \label{eq:gamma}
\end{align}
where the operator $\hat{ \mathrm{a} }_{ \bm{p} }$ is defined as $\hat{ \mathrm{a} }_{ \bm{p} } = |0 \rangle _{\mathrm{s}} \langle \bm{p} |$. 
The coefficient $\gamma(p)$ is a constant independent of $\bm{p}$ because it is Lorentz-invariant under $p^\mu \to (\Lambda p)^\mu$. 
Explicitly, it is computed as
\begin{align}
    \gamma(p) = \frac{\lambda^2}{m_\text{s}} \sqrt{m_\text{s}^2 - 4m_\text{E}^2} \ \theta\left(m_\text{s}^2 - 4 m_\text{E}^2 \right). 
    \label{eq:decay_rate}
\end{align}
Note that the $\theta\left(m_\text{s}^2 - 4 m_\text{E}^2 \right)$ is step function. Therefore, we finally obtain the following GKSL generator:
\begin{align}
   \mathcal{L}_{\phi}[ \rho^\text{in}_{\mathrm{s}}]
    = \frac{\lambda^2}{m_\text{s}} \sqrt{m_\text{s}^2 - 4m_\text{E}^2} \ \theta\left(m_\text{s}^2 - 4 m_\text{E}^2 \right)
     \int \frac{d^3 p \, d^3 \bar{p}}{\sqrt{ \omega_{\bm{p}}\,\omega_{\bar{\bm{p}}} }} \delta^4(p - \bar{p} )
    \left[ \hat{ \mathrm{a} }_{ \bm{p} } \rho^\text{in}_{ \mathrm{s} }\hat{ \mathrm{a} }^{\dagger}_{\bm{ \bar{p} }} - \frac{1}{2} \left\{ \hat{ \mathrm{a} }^{\dagger}_{\bm{ \bar{p} } } \hat{ \mathrm{a} }_{ \bm{p} }, \rho^\text{in}_{ \mathrm{s} } \right\} \right]
    \label{eq:L_decay2}
\end{align}
Here, we enumerate the three properties of this generator in TABLE \ref{tab:property_decay}. 
Firstly, the step function $\theta\left(m_\text{s}^2 - 4 m_\text{E}^2 \right)$ respects the fact that the particle decay is forbidden if $m_\text{s}< 2 m_\text{E}$ by the law of energy conservation.
Secondly, the generator \eqref{eq:L_decay2} describes the inevitable decay of particle for the infinite time. 
Indeed, Eq.\eqref{eq:L_decay2} diverges because of the delta function $\delta^4(p-\bar{p})=\delta(0)\delta^3(\bm{p}-\bar{\bm{p}})$. The divergence of the energy delta function $\delta(0)$ comes from taking the infinite time limit. This suggests that the decay process is dominant for a long timescale.
If one wants to have a finite form, the cutoff of evolving time should be introduced.
Finally, the generator has Poincar\'e symmetry if we consider the infinite time of this process. The Poincar\'e symmetry is defined by
\begin{align}
    \hat{U} \mathcal{L}_{\phi}[\rho^\text{in}_{\mathrm{s}}] \hat{U}^{\dagger}
    = \mathcal{L}_{\phi} \left[ \hat{U}\rho^\text{in}_{\mathrm{s}}\hat{U}^{\dagger}\right],
    \label{eq:Symmetry_decay}
\end{align}
where $\hat{U}=\hat{U}(\Lambda,a)$ is the unitary representation of Poincar\'e transformation.
This definition means that the GKSL generator is invariant under Poincar\'e transformations \cite{Kaito_2024}.
We can check the symmetric property in Appendix \ref{app:Confirm_symmetry} through the following relation:
\begin{align}
    \hat{U} \hat{ \mathrm{a} }^{\dagger}_{\bm{p}} \hat{U}^{\dagger}
    = \sqrt{ \frac{\omega_{\bm{p}_\Lambda}}{\omega_{\bm{p}}}}
    e^{ -i(\Lambda p)^\mu a_\mu }
    \hat{ \mathrm{a} }^{\dagger}_{ \bm{p}_\Lambda }
    \label{eq:Poincare_trans}
\end{align}
with $(\bm{p}_\Lambda)^i=(\Lambda p)^i$. 
As mentioned above, the generator of Eq.\eqref{eq:L_decay2} has the divergence due to the infinite time and implies the inevitable decay.
So we should consider that the Poincar\'{e} symmetry emerges for the asymptotic evolution of OQS from an in-state $\rho^\text{in}_\text{s}$ to the vacuum state $\rho^\text{out}_\text{s}=|0\rangle_\text{s} \langle 0|$.
If the cutoff of evolving time is introduced, the Poincar\'e symmetry is broken. 
\begin{table}[H]
    \centering
    \begin{tabular}{|c|}
    \hline
     The properties of $\mathcal{L}_{\phi}\text[\rho^\text{in}_\text{s}]$ \\ \hline
     \begin{minipage}{0.8\textwidth}
         \vspace{5pt}
         \begin{enumerate}
             \item This generator asymptotically has Poincar\'e symmetry.
             \item This generator diverges and this reflects the inevitable decay of particle.
             \item The decay of particle is forbidden if $m_\text{s} < 2m_\text{E}$.
         \end{enumerate}
         \vspace{3pt}
     \end{minipage} \\ \hline
    \end{tabular}
    \caption{Properties of $\mathcal{L}_{\phi}[\rho^\text{in}_\text{s}]$}
    \label{tab:property_decay}
\end{table}

\subsection{Pair annihilation and $2\rightarrow 2$ scattering of scalar particles}
\label{sec:Pair_annihilation}
\begin{figure}[H]
  \begin{minipage}[b]{0.5\textwidth}
    \centering
    \begin{tikzpicture}
    \definecolor{darkgreen}{RGB}{0,100,0}
    \begin{feynman}
      \vertex (a) at (-2,-2) {$\phi$};
      \vertex (b) at (-2,2) {$\chi$};
      \vertex (c) at (2,-2) {$\phi$};
      \vertex (d) at (2,2) {$\chi$};
      \vertex (i1) at (-1,0);
      \vertex (i2) at (1,0);
      \vertex (j) at (0,0) [label=below:{$\chi$}];

      \diagram* {
        (a) -- [scalar, momentum=\(p_1\)] (i1) -- [momentum=\(k_1\)](b),
        (i1) -- (i2),
        (i2) -- [momentum={[arrow distance = -3.0mm, label distance=-15pt]\(k_2\)}](d),
        (c) -- [scalar, momentum={[arrow distance=-3.0mm, label distance=-15pt] \(p_2\)}] (i2),
      };
      \fill (i1) circle (2pt);
      \fill (i2) circle (2pt);
    \end{feynman}
  \end{tikzpicture}
  \end{minipage}
  \begin{minipage}[b]{0.5\textwidth}
    \centering
    \begin{tikzpicture}
    \definecolor{darkgreen}{RGB}{0,100,0}
    \begin{feynman}
      \vertex (a) at (-2,-2) {$\phi$};
      \vertex (b) at (-2,2) {$\chi$};
      \vertex (c) at (2,-2) {$\phi$};
      \vertex (d) at (2,2) {$\chi$};
      \vertex (i1) at (-1,0);
      \vertex (i2) at (1,0);
      \vertex (j) at (0,0)[label=below:{$\chi$}];
      \vertex (k1) at (-0.5,0.99);
      \vertex (k2) at (0.5,0.99);
      \vertex (l1) at (-1.0,1.0);
      \vertex (l2) at (1.0,1.0);

      \diagram* {
        (a) -- [scalar, momentum=\(p_1\)] (i1),
        (i1) -- (i2),
        (c) -- [scalar, momentum={[arrow distance=-3.0mm, label distance=-15pt] \(p_2\)}] (i2),
        (i1) -- (k2) --[momentum={[arrow distance=-3.0mm, label distance=-15pt] \(k_2\)}] (d),
        (i2) -- (k1) -- [momentum=\(k_1\)] (b),
      };
      \fill (i1) circle (2pt);
      \fill (i2) circle (2pt);
    \end{feynman}
  \end{tikzpicture}
  \end{minipage}
  \caption{The diagrams representing the pair annihilation of $\phi$ particles}
  \label{fig:Pair}
\end{figure}
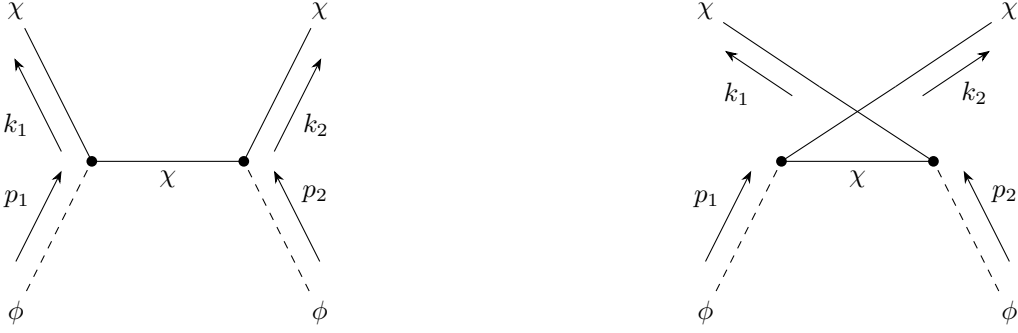

Following the same procedure as in the previous subsection, we will derive the generator for pair annihilation and $2\rightarrow 2$ scattering processes. 
We here assume that the particle $\phi$ is stable, that is, the relation $m_\text{s} < 2m_\text{E}$ holds and the decay process of particle $\phi$ does not occur. 
We first derive the Lindblad operator $\hat{T}_\beta$ given by the pair annihilation process ($\phi\phi \rightarrow \chi \chi$) denoted in Fig.\ref{fig:Pair}.
The difference with Sec. \ref{sec:Decay_process} is that we need to expand the S-operator \eqref{eq:S} up to the second order of $\lambda$. 
The S-operator up to $O(\lambda^2)$ is 
\begin{align}
    \hat{S} \approx \hat{ \mathbb{I} } + i \lambda \int d^4 x \ \hat{\phi}(t,\bm{x}) \otimes \hat{\chi}^2(t,\bm{x}) - \frac{\lambda^2}{2!} \int d^4 x d^4 y \ \mathrm{T} \left[ \hat{\phi}(x)\hat{\phi}(y) \otimes \hat{\chi}^2(x) \hat{\chi}^2(y) \right],
    \label{eq:S_pair}
\end{align}
so we read out the operator $\hat{T}$ as
\begin{align}
    \hat{T} =
    \lambda \int d^4 x \ \hat{\phi}(t,\bm{x}) \otimes \hat{\chi}^2(t,\bm{x}) + \frac{i\lambda^2}{2!} \int d^4 x  d^4 y \ \mathrm{T} \left[ \hat{\phi}(x)\hat{\phi}(y) \otimes \hat{\chi}^2(x) \hat{\chi}^2(y) \right].
    \label{eq:T_pair}
\end{align}
Respecting the Feynmann diagram in Fig.\ref{fig:Pair}, we have the following Lindblad operator $\hat{T}_{\beta}=\hat{T}_{\bm{k}_1 \bm{k}_2}$, 
\begin{align}
    \hat{T}_{\bm{k}_1 \bm{k}_2} =  \int d^3 p_1 d^3 p_2 \left[ {}_{\mathrm{E}}\langle \bm{k}_1, \bm{k}_2 | {}_{\mathrm{s}}\langle 0 | \  \hat{T} \ | \bm{p}_1,\bm{p}_2 \rangle_{\mathrm{s}} | 0 \rangle _{\mathrm{E}}  \right] |0 \rangle _{\mathrm{s}} \langle \bm{p}_1, \bm{p}_2 |.
    \label{eq:T^s_beta3}
\end{align}
The scattering amplitude 
${}_{\mathrm{E}}\langle \bm{k}_1, \bm{k}_2 | {}_{\mathrm{s}}\langle 0 | \ i \hat{T} \ | \bm{p}_1,\bm{p}_2 \rangle_{\mathrm{s}} | 0 \rangle _{\mathrm{E}}$ includes the Feynman diagrams of the first and second order of $\lambda$, but the diagram with $O(\lambda)$ does not appear because one vertex can not connect the two external lines of $\phi$ particle with the two external lines of $\chi$ particle. 
Thus, we only need to consider the diagrams given in Fig. \ref{fig:Pair}. These yield the following scattering amplitude, 
\begin{align}
    {}_{\mathrm{E}}\langle \bm{k}_1, \bm{k}_2 | {}_{\mathrm{s}}\langle 0 | \ i \hat{T} \ | \bm{p}_1,\bm{p}_2 \rangle_{\mathrm{s}} | 0 \rangle _{\mathrm{E}}
    &= - \frac{2\lambda^2}{(2\pi)^2} \frac{\delta^4(k_1 + k_2 - p_1 - p_2)}{ \sqrt{E_{\bm{k}_1} E_{\bm{k}_2} \omega_{\bm{p}_1} \omega_{\bm{p}_2} } }
    \left[ D_{ \mathrm{F} }(p_2 - k_1) + D_{ \mathrm{F} } (p_2 - k_2) \right],
    \label{eq:S_pair2}
\end{align}
where $D_{ \mathrm{F} }(q)$ is the Feynman propagator in momentum space, 
\begin{align}
    D_{ \mathrm{F} }(q) = \frac{1}{q^2 + m_\text{E}^2 - i \epsilon}.
    \label{eq:Propagator}
\end{align}
Substituting Eq.\eqref{eq:S_pair2} for Eq.\eqref{eq:T^s_beta3}, we have the Lindblad operator 
\begin{align}
    \hat{T}_{\bm{k}_1 \bm{k}_2} =  \frac{2i\lambda^2}{(2\pi)^2\sqrt{ E_{\bm{k}_1} E_{\bm{k}_2} }}\int \frac{d^3 p_1 d^3 p_2}{\sqrt{\omega_{\bm{p}_1} \omega_{\bm{p}_2}}}  \delta^4(k_1 + k_2 - p_1 - p_2)
    \left[ D_{ \mathrm{F} }(p_2 - k_1) + D_{ \mathrm{F} } (p_2 - k_2) \right]|0 \rangle _{\mathrm{s}} \langle \bm{p}_1, \bm{p}_2 |. 
    \label{eq:T^s_beta4}
\end{align}
This gives the second term of the GKSL generator in \eqref{eq:L_scatt}, which leads to the non-unitary evolution of $\phi$ particles.
Explicitly, it has the following form, 
\begin{align}
    &\text{The 2nd term of Eq.\eqref{eq:L_scatt}}
    \notag
    \\
    &  \quad =\frac{4\lambda^4}{(2\pi)^4} \int \frac{d^3 \bar{p}_1 d^3 \bar{p}_2 d^3 p_1 d^3 p_2}{\sqrt{\omega_{\bar{\bm{p}}_1}\omega_{\bar{\bm{p}}_2}\omega_{\bm{p}_1}\omega_{\bm{p}_2}}} 
    \ \delta^4 (\bar{p}_1 + \bar{p}_2 - p_1 - p_2 ) \gamma(p_1,p_2,\bar{p}_2) 
    \left[ 
    \hat{\mathrm{a}}_{\bm{p}_1 \bm{p}_2} 
    \rho^\text{in}_{\mathrm{s}}
    \hat{\mathrm{a}}^{\dagger}_{\bm{\bar{p}}_1 \bm{\bar{p}}_2}
    - \frac{1}{2} \left\{ 
    \hat{\mathrm{a}}^{\dagger}_{\bm{\bar{p}}_1 \bm{\bar{p}}_2}
    \hat{\mathrm{a}}_{\bm{p}_1 \bm{p}_2},
    \rho^\text{in}_{\mathrm{s}}
    \right\}
    \right],
    \label{eq:2nd_Lpair}
\end{align}
where $\hat{\mathrm{a}}_{\bm{p}\,\bm{p}'}=|0\rangle_\text{s} \langle \bm{p}, \bm{p}'|$  and 
the $\gamma(p_1,p_2,\bar{p}_2)$ is 
\begin{align}
    \gamma(p_1,p_2,\bar{p}_2) &=
    \int \frac{d^3 k_1 d^3 k_2}{E_{\bm{k}_1} E_{\bm{k}_2}} \
    \delta^4(p_1 + p_2 - k_1 - k_2) 
    \left[ 
    D_{ \mathrm{F} }(p_2 - k_2) + D_{ \mathrm{F} }(p_2 - k_1) 
    \right]
    \left[
    D^\ast _{ \mathrm{F} }(\bar{p}_2 - k_2) + D^\ast _{ \mathrm{F} }(\bar{p}_2 - k_1)
    \right].
    \label{eq:gamma_pair}
\end{align}
This is invariant under Lorentz transformations $p^\mu_i \rightarrow (\Lambda p_i)^\mu$ and $\bar{p}^\mu_j \rightarrow (\Lambda \bar{p}_j)^\mu$ and characterizes the pair annihilation process. 
Note that the loop diagrams of the second-order of $\lambda$ exist other than the Feynman diagrams in Fig.\ref{fig:Pair}, but it does not matter since they are ignored by the redefinition of S-operator and by the mass renormalization.
At this stage, we have completed the analysis on the pair annihilation process and have found that it yields the Lindblad operator $\hat{T}_\beta=\hat{T}_{\bm{k}_1 \bm{k}_2}$.  
However, we have not computed $\text{Re}\hat{T}_0$ that gives the first term in the GKSL generator. 
In the following, we will see that $\text{Re}\hat{T}_0$ is computed from the $2\rightarrow 2$ scattering of particles.

Since the Lindblad operator $\hat{T}_\beta=\hat{T}_{\bm{k}_1 \bm{k}_2}$ has $O(\lambda^2)$, 
the second term of the GKSL generator has $O(\lambda^4)$ as shown in Eq.\eqref{eq:2nd_Lpair}.
Hence, the first term of the generator with $\text{Re}\hat{T}_0$ should be evaluated up to $O(\lambda^4)$.  
In order to calculate $\text{Re}\hat{T}_0$ given by $\hat{T}_0={}_\text{E} \langle 0|\hat{T}|0\rangle_\text{E}$, we should focus on the process where the environment does not excite. 
Here, we take the following form of $\hat{T}_0 $:
\begin{align}
    \hat{T}_0 
    = \int d^3 \bar{p}_1  d^3 \bar{p}_2  d^3 p_1 d^3 p_2   \left[ {}_{\mathrm{E}}\langle 0 | {}_\mathrm{s} \langle \bm{\bar{p}}_1, \bm{\bar{p}}_2 | \ \hat{T} \ | \bm{p}_1,\bm{p}_2 \rangle_{\mathrm{s}}| 0 \rangle _{\mathrm{E}} \right] 
    | \bm{\bar{p}}_1, \bm{\bar{p}}_2 \rangle _{\mathrm{s}} \langle \bm{p}_1, \bm{p}_2 |.
    \label{eq:T^s_0-pair}
\end{align}
The amplitude $ {}_{\mathrm{E}}\langle 0 | {}_{\mathrm{s}}\langle \bm{\bar{p}}_1, \bm{\bar{p}}_2 | \ i \hat{T} \ | \bm{p}_1,\bm{p}_2 \rangle_{\mathrm{s}} | 0 \rangle _{\mathrm{E}}$ is that of the $2\rightarrow 2$ scattering of $\phi$ particles $(\phi\phi \rightarrow \phi \phi)$. 
Up to the fourth order of $\lambda$, this process is dominant.
We should note that the diagrams with $O(\lambda)$ and $O(\lambda^3)$ do not give the amplitude $ {}_{\mathrm{E}}\langle 0 | {}_{\mathrm{s}}\langle \bm{\bar{p}}_1, \bm{\bar{p}}_2 | \ i \hat{T} \ | \bm{p}_1,\bm{p}_2 \rangle_{\mathrm{s}} | 0 \rangle _{\mathrm{E}}$. 
This is because we cannot connect the four external lines of $\phi$ particles by adequately using one vertex or three vertices with the three-point interaction $\lambda \phi \chi^2$.
Also, the diagrams with $O(\lambda^2)$ do not give the amplitude because they can be removed by the redefinition of S-operator and the mass renormalization of $\phi$ field. Furthermore, the most of the diagrams with $O(\lambda^4)$ can be removed. These details are presented in Appendix \ref{app:explanation of ReT_0 in 2-2}.
\begin{figure}[H]
  \begin{minipage}[t]{0.33\textwidth}
    \centering
    \begin{tikzpicture}
    \definecolor{darkgreen}{RGB}{0,100,0}
    \begin{feynman}
      \vertex (a) at (-2,-2) {$\phi$};
      \vertex (b) at (-2,2) {$\phi$};
      \vertex (c) at (2,-2) {$\phi$};
      \vertex (d) at (2,2) {$\phi$};
      \vertex (i1) at (-1,1);
      \vertex (i2) at (1,1);
      \vertex (i3) at (-1,-1);
      \vertex (i4) at (1,-1);
      \vertex (n1) at (-1,0) [label=left:{$\chi$}];
      \vertex (n2) at (0,1)  [label=above:{$\chi$}];
      \vertex (n3) at (1,0)  [label=right:{$\chi$}];
      \vertex (n4) at (0,-1) [label=below:{$\chi$}];
      \diagram* {
        (a) -- [scalar, momentum=\(p_1\)] (i3) -- (i4),
        (c) -- [scalar, momentum={[arrow distance=-3.0mm, label distance=-15pt] \(p_2\)}] (i4) -- (i2) -- [scalar, momentum={[arrow distance=-3.0mm, label distance=-15pt] \(\bar{p}_2\)}] (d),
        (i1) -- (i2),
        (i3) -- (i1),
        (i1) -- [scalar,momentum=\(\bar{p}_1\)] (b),
      };
      \fill (i1) circle (2pt);
      \fill (i2) circle (2pt);
      \fill (i3) circle (2pt);
      \fill (i4) circle (2pt);
      \node[circle,inner sep=2pt] (A) at (2.8,2) {};
      \node[circle,inner sep=2pt] (B) at (2.8,-2) {};
    \end{feynman}
  \end{tikzpicture}
  \end{minipage}
  \begin{minipage}[t]{0.33\textwidth}
    \centering
    \begin{tikzpicture}
    \definecolor{darkgreen}{RGB}{0,100,0}
    \begin{feynman}
      \vertex (a) at (-2,-2) {$\phi$};
      \vertex (b) at (-2,2) {$\phi$};
      \vertex (c) at (2,-2) {$\phi$};
      \vertex (d) at (2,2) {$\phi$};
      \vertex (i1) at (-1,1) ;
      \vertex (i2) at (1,1) ;
      \vertex (i3) at (-1,-1) ;
      \vertex (i4) at (1,-1) ;
      \vertex (n1) at (-1,0) [label=left:{$\chi$}];
      \vertex (n2) at (-0.4,0.5)  [label=above:{$\chi$}];
      \vertex (n3) at (1,0)  [label=right:{$\chi$}];
      \vertex (n4) at (0.4,0.5) [label=above:{$\chi$}];
      \diagram* {
        (a) -- [scalar, momentum=\(p_1\)] (i3) -- (i2),
        (c) -- [scalar, momentum={[arrow distance=-3.0mm, label distance=-15pt] \(p_2\)}] (i4) -- (i2) -- [scalar, momentum={[arrow distance=-3.0mm, label distance=-15pt] \(\bar{p}_2\)}] (d),
        (i3) -- (i1),
        (i1) -- (i4),
        (i1) -- [scalar,momentum=\(\bar{p}_1\)] (b),
      };
      \fill (i1) circle (2pt);
      \fill (i2) circle (2pt);
      \fill (i3) circle (2pt);
      \fill (i4) circle (2pt);
      \node[circle,inner sep=2pt] (A) at (3,2) {};
      \node[circle,inner sep=2pt] (B) at (3,-2) {};
    \end{feynman}
  \end{tikzpicture}
  \end{minipage}
  \begin{minipage}[t]{0.33\textwidth}
    \centering
    \begin{tikzpicture}
    \definecolor{darkgreen}{RGB}{0,100,0}
    \begin{feynman}
      \vertex (a) at (-2,-2) {$\phi$};
      \vertex (b) at (-2,2) {$\phi$};
      \vertex (c) at (2,-2) {$\phi$};
      \vertex (d) at (2,2) {$\phi$};
      \vertex (i1) at (-1,1) ;
      \vertex (i2) at (1,1) ;
      \vertex (i3) at (-1,-1) ;
      \vertex (i4) at (1,-1) ;
      \vertex (n1) at (-0.6,0.5) [label=below:{$\chi$}];
      \vertex (n2) at (0,1)  [label=above:{$\chi$}];
      \vertex (n3) at (0,-1)  [label=below:{$\chi$}];
      \vertex (n4) at (0.6,0.5) [label=below:{$\chi$}];
      \diagram* {
        (a) -- [scalar, momentum=\(p_1\)] (i3) -- (i2) -- [scalar, momentum={[arrow distance=-3.0mm, label distance=-15pt] \(\bar{p}_2\)}] (d),
        (c) -- [scalar, momentum={[arrow distance=-3.0mm, label distance=-15pt] \(p_2\)}] (i4) -- (i1),
        (i2) -- (i1),
        (i3) -- (i4),
        (i1) -- [scalar,momentum=\(\bar{p}_1\)] (b),
      };
      \fill (i1) circle (2pt);
      \fill (i2) circle (2pt);
      \fill (i3) circle (2pt);
      \fill (i4) circle (2pt);
    \end{feynman}
  \end{tikzpicture}
  \end{minipage}
  \caption{The 4th-order diagrams we should compute}
  \label{fig:4th-order diagram}
\end{figure}
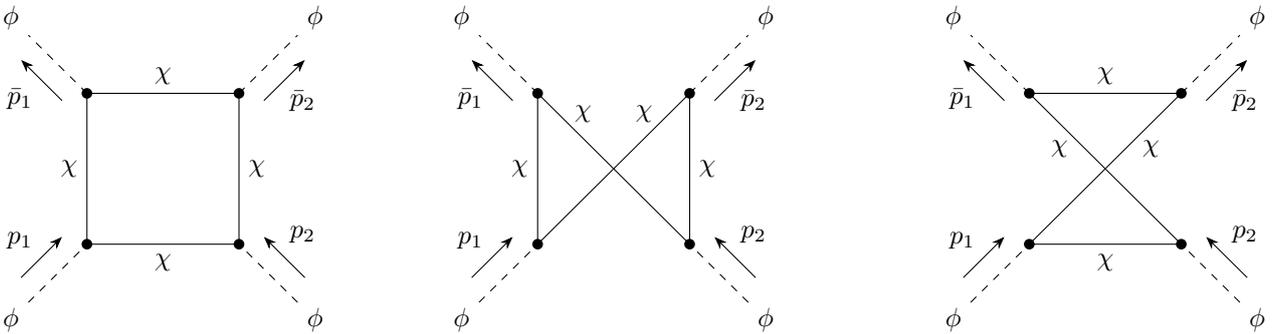

In the end, the diagrams depicted in Fig.\ref{fig:4th-order diagram} provide the amplitude ${}_{\mathrm{E}}\langle 0 |{}_{\mathrm{s}}\langle \bm{\bar{p}}_1, \bm{\bar{p}}_2 | \ i \hat{T} \ | \bm{p}_1,\bm{p}_2 \rangle_{\mathrm{s}} | 0 \rangle _{\mathrm{E}}$ at the fourth order, 
\begin{align}
  &{}_{\mathrm{E}}\langle 0 | {}_{\mathrm{s}}\langle \bm{\bar{p}}_1, \bm{\bar{p}}_2 | \ i \hat{T} \ | \bm{p}_1,\bm{p}_2 \rangle_{\mathrm{s}} | 0 \rangle _{\mathrm{E}} 
  \notag
  \\
  & \quad 
  = \frac{ (2\pi)^{ - \frac{3}{2} } }{ \sqrt{ 2\omega_{\bm{p}_1} } } \cdot \frac{ (2\pi)^{ - \frac{3}{2} } }{ \sqrt{ 2\omega_{\bm{p}_2} } } \cdot \frac{ (2\pi)^{ - \frac{3}{2} } }{ \sqrt{ 2\omega_{\bar{\bm{p}}_1} } } \cdot \frac{ (2\pi)^{ - \frac{3}{2} } }{ \sqrt{ 2\omega_{\bar{\bm{p}}_2} } } \cdot (-i\lambda)^4
  \int d^4 q_1 d^4 q_2 d^4 \bar{q}_1 d^4 \bar{q}_2
  \frac{ D_\mathrm{F}(q_1)  }{(2\pi)^4} \cdot 
  \frac{ D_\mathrm{F}(q_2)  }{(2\pi)^4} \cdot
  \frac{ D_\mathrm{F}(\bar{q}_1)  }{(2\pi)^4} \cdot 
  \frac{ D_\mathrm{F}(\bar{q}_2)  }{(2\pi)^4} \notag \\
  & \quad 
  \times (2\pi)^{16} \left[ 
  \delta^4(q_1 + q_2 - p_1) 
  \delta^4(\bar{q}_2 - q_2 - p_2) 
  \delta^4(\bar{p}_1 + \bar{q}_1 - q_1)
  \delta^4(\bar{p}_2 - \bar{q}_1 - \bar{q}_2)
  \right. \notag \\
  & \qquad \qquad \quad \left. + 
  \delta^4(q_1 + q_2 - p_1) 
  \delta^4(\bar{q}_1 - \bar{q}_2 - p_2) 
  \delta^4(\bar{p}_1 - q_1 - \bar{q}_2)
  \delta^4(\bar{p}_2 - \bar{q}_1 - q_2)
  \right. \notag \\
  & \qquad \qquad \quad \left. + 
  \delta^4(q_1 + q_2 - p_1) 
  \delta^4(\bar{q}_2 - q_1 - p_2) 
  \delta^4(\bar{p}_1 + \bar{q}_1 - \bar{q}_2)
  \delta^4(\bar{p}_2 - \bar{q}_1 - q_2)
  \right] \notag \\
  & \quad = \frac{ 2 \lambda^4 }{ (2\pi)^6 \sqrt{ \omega_{\bm{p}_1} \omega_{\bm{p}_2} \omega_{\bar{\bm{p}}_1} \omega_{\bar{\bm{p}}_2} } } \ \mathcal{A}( p_1, p_2, \bar{p}_1, \bar{p}_2 ) \
  \delta^4 (\bar{p}_1 + \bar{p}_2 - p_1 - p_2 ),
  \label{eq:S_4th}
\end{align}
where $\mathcal{A}(p_1,p_2,\bar{p}_1,\bar{p}_2)$ is defined as
\begin{align}
&\mathcal{A}( p_1, p_2, \bar{p}_1, \bar{p}_2 )
\notag 
\\
&\quad =\int d^4 q \left[ D_{\mathrm{F}}(q) D_{\mathrm{F}}(q - p_1) D_{\mathrm{F}}(q - \bar{p}_1) D_{\mathrm{F}}(q + \bar{p}_2 - p_1)  + D_{\mathrm{F}}(q) D_{\mathrm{F}}(q - p_1) D_{\mathrm{F}}(q - \bar{p}_1) D_{\mathrm{F}}(q - p_1 - p_2) \right. \notag \\
&\qquad  \left. + D_{\mathrm{F}}(q) D_{\mathrm{F}}(q - p_1) D_{\mathrm{F}}(q + p_2) D_{\mathrm{F}}(q + \bar{p}_2 - p_1)
  \right]. 
  \label{eq:A_def}
\end{align}
In proceeding with the calculation of $\mathcal{A}( p_1, p_2, \bar{p}_1, \bar{p}_2 )$, the Mandelstam variables $s = (p_1 + p_2)^2$, $t =  \ (p_1 - \bar{p}_1)^2$, $u = (p_1 - \bar{p}_2)^2$ and the Feynman parameter integral are useful \cite{Peskin_1995, BBJ_2007}. 
The coefficient $\mathcal{A}( p_1, p_2, \bar{p}_1, \bar{p}_2 )$ is rewritten as
\begin{align}
    \mathcal{A}(s, t, u)
    &= \frac{i}{(4\pi)^2}  \int^{1}_{0} dz_1 dz_2 dz_3 dz_4 \ 
    \delta(1 - z_1 - z_2 - z_3 - z_4) 
    \left[ 
    \frac{1}{ (M^2_{1} - i\epsilon)^2 } 
    + \frac{1}{ (M^2_{2} - i\epsilon)^2} 
    + \frac{1}{ (M^2_{3} - i\epsilon)^2}
    \right],
    \label{eq:A_new}
\end{align}
where the integral domain of $z_i$ is $0\leq z_i \leq 1$, and $M^2_1, M^2_2, M^2_3$ are defined respectively as
\begin{align}
   M^2_1 &= m_\text{E}^2 - (z_1 + z_4)(z_2 + z_3)m_\text{s}^2 + z_2 z_3 t + z_1 z_4 u,
   \label{eq:M1_def}
   \\
   M^2_2 &= m_\text{E}^2 - (z_1 + z_4)(z_2 + z_3)m_\text{s}^2 + z_2 z_3 t + z_1 z_4 s,
   \label{eq:M2_def}
   \\
   M^2_3 &= m_\text{E}^2 - (z_1 + z_4)(z_2 + z_3)m_\text{s}^2 + z_2 z_3 s + z_1 z_4 u.
   \label{eq:M3_def}
\end{align}
These are written as a function of the Mandelstam variables $s$, $t$, and $u$. 
We derive the expresstion of $\mathcal{A}$ in Appendix \ref{app:Def_Anew}
We have obtained the expression of $\hat{T}_0$, but we really want to get $\mathrm{Re} \hat{T}_0$. By the definition, we find the following expression of $\mathrm{Re} \hat{T}_0$:
\begin{align}
    \mathrm{Re} \hat{T}_0 
    &= \frac{1}{2} ( \hat{T}_0 + \hat{T}^{\dagger}_0 ) \notag \\
    &= - \frac{i}{2} \int d^3 \bar{p}_1 \int d^3 \bar{p}_2 \int d^3 p_1 \int d^3 p_2 \ \frac{ 8 \lambda^4 }{ (2\pi)^6 \sqrt{ 2\omega_{\bm{p}_1} 2\omega_{\bm{p}_2} 2\omega_{\bar{\bm{p}}_1} 2\omega_{\bar{\bm{p}}_2} } } \delta^4 (\bar{p}_1 + \bar{p}_2 - p_1 - p_2 ) \notag \\
    & \quad \times 
    \Big[ 
    \mathcal{A}(s,t,u) | \bm{\bar{p}}_1, \bm{\bar{p}}_2 \rangle _{\mathrm{s}} \langle \bm{p}_1, \bm{p}_2 | 
    - 
    \mathcal{A}^{\ast}(s,t,u) | \bm{p}_1, \bm{p}_2 \rangle _{\mathrm{s}} \langle \bm{\bar{p}}_1, \bm{\bar{p}}_2 | 
    \Big] \notag \\
    &= \frac{2\lambda^4}{(2\pi)^6}  
    \int \frac{d^3 \bar{p}_1 d^3 \bar{p}_2 d^3 p_1 d^3 p_2}{\sqrt{\omega_{\bar{\bm{p}}_1}\omega_{\bar{\bm{p}}_2}\omega_{\bm{p}_1}\omega_{\bm{p}_2}}}
    \ \mathrm{Im}\mathcal{A}(s,t,u) \
    \delta^4 (\bar{p}_1 + \bar{p}_2 - p_1 - p_2 ) 
    | \bm{\bar{p}}_1, \bm{\bar{p}}_2 \rangle _{\mathrm{s}} \langle \bm{p}_1, \bm{p}_2 |.
    \label{Real_T^s0}
\end{align}
Hence we have the first term of the GKSL generator,
\begin{align}
    \text{The 1st term of Eq.\eqref{eq:L_scatt}}
     = i \frac{2\lambda^4}{(2\pi)^6} \int \frac{d^3 \bar{p}_1 d^3 \bar{p}_2 d^3 p_1 d^3 p_2}{\sqrt{\omega_{\bar{\bm{p}}_1}\omega_{\bar{\bm{p}}_2}\omega_{\bm{p}_1}\omega_{\bm{p}_2}}}
     \ \delta^4 (\bar{p}_1 + \bar{p}_2 - p_1 - p_2 ) \mathrm{Im}\mathcal{A}(s,t,u) 
     \left[ 
     \hat{\mathrm{a}}^{\dagger}_{\bm{\bar{p}}_1 \bm{\bar{p}}_2}
     \hat{\mathrm{a}}_{\bm{p}_1 \bm{p}_2},
     \rho^\text{in}_{\mathrm{s}}
     \right] .
    \label{eq:1st_Lpair}
\end{align}
Let $\mathcal{L}_{\phi\phi}$ denote the GKSL generator associated with the pair annihilation and the $2\to 2$ scattering processes. 
Combining the results \eqref{eq:2nd_Lpair} and \eqref{eq:1st_Lpair}, we arrive at the following GKSL generator:
\begin{align}
    &\mathcal{L}_{\phi\phi}[\rho^\text{in}_{\mathrm{s}}] 
    = \mathcal{H}[\rho^\text{in}_{\mathrm{s}}] + \mathcal{D}[\rho^\text{in}_{\mathrm{s}}], 
    \label{eq:L_pair}
    \\
    & \mathcal{H}[\rho^\text{in}_{\mathrm{s}}]
    = i \frac{2\lambda^4}{(2\pi)^6} \int \frac{d^3 \bar{p}_1 d^3 \bar{p}_2 d^3 p_1 d^3 p_2}{\sqrt{\omega_{\bar{\bm{p}}_1}\omega_{\bar{\bm{p}}_2}\omega_{\bm{p}_1}\omega_{\bm{p}_2}}}
     \ \delta^4 (\bar{p}_1 + \bar{p}_2 - p_1 - p_2 ) \mathrm{Im}\mathcal{A}(s,t,u) 
     \left[ 
     \hat{\mathrm{a}}^{\dagger}_{\bm{\bar{p}}_1 \bm{\bar{p}}_2}
     \hat{\mathrm{a}}_{\bm{p}_1 \bm{p}_2},
     \rho^\text{in}_{\mathrm{s}}
     \right] 
     \label{eq:H_pair}
     \\
    & \mathcal{D} [\rho^\text{in}_{\mathrm{s}}]
    = \frac{4\lambda^4}{(2\pi)^4} \int \frac{d^3 \bar{p}_1 d^3 \bar{p}_2 d^3 p_1 d^3 p_2}{\sqrt{\omega_{\bar{\bm{p}}_1}\omega_{\bar{\bm{p}}_2}\omega_{\bm{p}_1}\omega_{\bm{p}_2}}} 
    \ \delta^4 (\bar{p}_1 + \bar{p}_2 - p_1 - p_2 ) \gamma(p_1,p_2,\bar{p}_2) 
    \left[ 
    \hat{\mathrm{a}}_{\bm{p}_1 \bm{p}_2} 
    \rho^\text{in}_{\mathrm{s}}
    \hat{\mathrm{a}}^{\dagger}_{\bm{\bar{p}}_1 \bm{\bar{p}}_2}
    - \frac{1}{2} \left\{ 
    \hat{\mathrm{a}}^{\dagger}_{\bm{\bar{p}}_1 \bm{\bar{p}}_2}
    \hat{\mathrm{a}}_{\bm{p}_1 \bm{p}_2},
    \rho^\text{in}_{\mathrm{s}}
    \right\}
    \right],
    \label{eq:D_pair}
\end{align}
where $\mathcal{H}$ and $\mathcal{D}$ comes from the $2\to2$ scattering ($\phi\phi \to \phi\phi)$ and the pair annhilation ($\phi\phi \to \chi\chi$).   

The features of the above GKSL generator are summarized in TABLE \ref{tab:property_pair}.
Let us examine each of these one by one.
Eq.\eqref{eq:D_pair} represents the dissipative behavior of scalar particles $\phi$ due to the pair annihilation process.
To check this fact, let us calculate the pair annihilation probability given as
\begin{align}
    {}_\text{s}\langle 0|\mathcal{L}_{\phi\phi}[\rho^\text{in}_{\mathrm{s}}]|0\rangle_\text{s} \notag 
    &= {}_\text{s}\langle 0|\mathcal{D}[\rho^\text{in}_{\mathrm{s}}]|0\rangle_\text{s} \notag \\
    &= \frac{4\lambda^4}{(2\pi)^4} \int \frac{d^3 \bar{p}_1 d^3 \bar{p}_2 d^3 p_1 d^3 p_2}{\sqrt{\omega_{\bar{\bm{p}}_1}\omega_{\bar{\bm{p}}_2}\omega_{\bm{p}_1}\omega_{\bm{p}_2}}} 
    \ \delta^4 (\bar{p}_1 + \bar{p}_2 - p_1 - p_2 ) \gamma(p_1,p_2,\bar{p}_2)
    {}_\text{s}\langle \bm{p}_1,\bm{p}_2|\rho^\text{in}_\text{s}|\bm{ \bar{p} }_1, \bm{ \bar{p} }_2\rangle_\text{s}, 
    \label{eq:probability_annihilation}
\end{align}
Hence we have the nontrivial transition probability from a two-particle in-state $\rho^\text{in}_\text{s}$ to the vacuum state $|0\rangle_\text{s}$, which is nothing but the pair annihilation process. 
Our formalism based on GKSL generator is possible to adopt the initial superposition state of $\phi$ particles. 
Let us imagine the superposition of the pair of incident $\phi$ particles with momenta $\bm{q}, -\bm{q}$ and $\bar{\bm{q}},-\bar{\bm{q}}$,
\begin{align}
    \rho^\text{in}_\text{s} = |\psi\rangle_\text{s}\langle\psi|, \qquad |\psi\rangle_\text{s} =  \frac{1}{\sqrt{N}} \Big( |\bm{q},-\bm{q}\rangle_\text{s} + e^{i\delta} |\bar{\bm{q}},-\bm{\bar{q}} \rangle_\text{s} \Big),
    \notag
\end{align}
where $N$ is a normalization and the parameter $\delta$ is a relative phase. 
We also assume that the momenta $\bm{q}$ and $\bar{\bm{q}}$ are orthogonal to each other, i.e., $\bm{q} \cdot \bar{\bm{q}}=0$, and have the same magnitude, $|\bm{q}|=|\bar{\bm{q}}|$. 
Then, the following probability is obtained:
\begin{align}
    &{}_\text{s}\langle 0|\mathcal{D}[\rho^\text{in}_{\mathrm{s}}]|0\rangle_\text{s} 
    \nonumber 
    \\
    & \quad 
    = \left( \frac{\lambda}{\pi} \right)^4 \frac{VT}{N} \frac{16 \pi }{-s} \sqrt{ \frac{s + 4m^2_\text{E}}{s} } \frac{\theta(-s - 4m^2_\text{E})}{ (s + 2m^2_\text{s})^2 }
    \left\{ \frac{16}{1-a} + \frac{8}{\sqrt{a}} \log\frac{1+\sqrt{a}}{1-\sqrt{a}} + \cos\delta \, \left[ \frac{11-a}{2\sqrt{a}} \log\frac{1+\sqrt{a}}{1-\sqrt{a}} \right] \right\}
    \label{eq:Prob_anni1}
\end{align}
where $a=(s + 4m^2_\text{s})(s + 4m^2_\text{E})/(s + 2m^2_\text{s})^2$ and we reguralized $\delta^4(0)$ by introducing the spacetime volume $\delta^4(0)=VT$. 
The probability is converted to a dimensionless quantity $m^2_\text{E}\sigma$ with
\begin{align}
    \sigma = \frac{N \times {}_\text{s}\langle 0|\mathcal{D}[\rho^\text{in}_{\mathrm{s}}]|0\rangle_\text{s}}{ m^6_\text{E} VT \, u }  , \qquad
    u = \frac{ \sqrt{ (q_1 \cdot q_2)^2 - m^4_\text{s} } }{ \omega_{q_1}\omega_{q_2} } = \frac{ 2\sqrt{s(s+4m^2_\text{s})} }{-s}, \notag
\end{align}
where $\sigma$ has the same dimension as cross section.
\begin{figure}[H]
    \centering
    \includegraphics[width=0.7\linewidth]{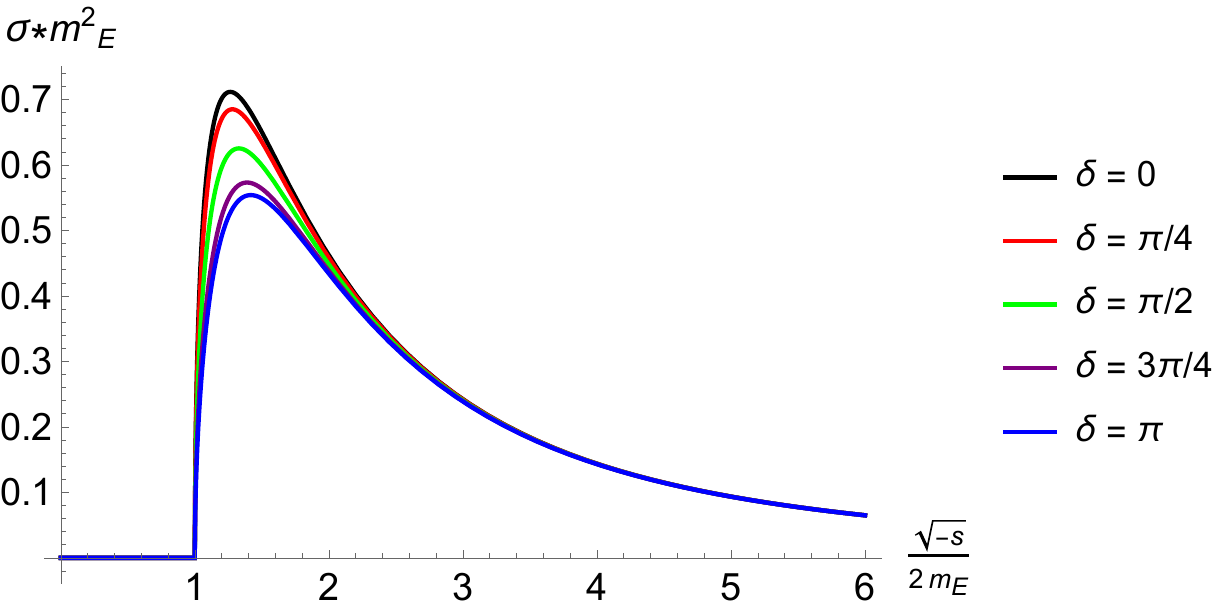}
    \caption{ 
    The behavior of $\sigma$ proportional to the pair annihilation probability. 
    The horizontal axis is the total energy of incident particles. 
    Note that both axes is normalized by the environment field mass $m_\text{E}$, and we plot the graph at the normalized coupling $\frac{\lambda}{2m_\text{E}} = 0.1$ and the normalized mass $\frac{m_\text{s}}{2m_\text{E}} = 0.01$. 
    The probability starts to increase at $\frac{ \sqrt{-s} }{2m_\text{E}} = 1$ and is dumped at high energy regime.
    Furthermore, the probability decreases as the relative phase $\delta$ increases.}
    \label{fig:Prob_0.01}
\end{figure}

As can be seen Fig.\ref{fig:Prob_0.01}, we can find three characteristic features. The first one is that the probability starts to rise at $\frac{ \sqrt{-s} }{2m_\text{E}} = 1$.
This is because that the energy conservation forbids the pair annihilation process at $\frac{ \sqrt{-s} }{2m_\text{E}} < 1$.
The second one is that the probability is dumped as the total energy increases.
This is thought to be because, as the energy increases, the cross-section experienced by the incident particles becomes smaller, which makes the interactions less likely to occur.
The last one is that the peak of the probability decreases as the pase difference $\delta$ increases. 
We consider this behavior in Sec.\ref{sec:Discussion}.
Our GKSL generator describes not only the standard pair annihilation but also the pair annihilation of incident superposed particles.

Eq.\eqref{eq:H_pair} reflects that the particles $\phi$ interact by exchanging the virtual particles $\chi$. This gives the unitary evolution and does not appear in the decay process of $\phi$ particle, which is consistent with the diagrams shown in Fig.\ref{fig:4th-order diagram}.
Comparing the two Eqs. \eqref{eq:D_pair} and \eqref{eq:H_pair}, we observe that the dynamics gets close to a unitary dynamics for $m^2_\text{s},|s|,|t|,|u| \ll m^2_\text{E}$. 
In fact, under this situation, the coefficient $\mathcal{A}(s,t,u)$  can be approximated as follow:
\begin{equation}
    \mathcal{A}(s,t,u) 
    \approx 
    \frac{3i}{(4\pi)^2} \int^1_0 dz_1 dz_2 dz_3 dz_4 \ \frac{\delta(1-z_1-z_2-z_3-z_4)}{m^4_\text{E}} 
    = \frac{3i}{(4\pi)^2 m^4_\text{E}} 
\end{equation}
On the other hand, the coefficient Eq.\ref{eq:gamma_pair} is rewritten by adopting the center-of-mass frame $\bm{p}_1+\bm{p}_2=0$, 
\begin{align}
    \gamma(p_1,p_2,\bar{p}_2) 
    &= \int \frac{d^3 k_1 d^3 k_2}{E_{\bm{k}_1} E_{\bm{k}_2}} \
    \delta^4(p_1 + p_2 - k_1 - k_2) 
    \left[ 
    D_{ \mathrm{F} }(p_2 - k_2) + D_{ \mathrm{F} }(p_2 - k_1) 
    \right]
    \left[
    D^\ast _{ \mathrm{F} }(\bar{p}_2 - k_2) + D^\ast _{ \mathrm{F} }(\bar{p}_2 - k_1)
    \right] \notag \\
    &= \frac{1}{2}\int d\Omega \sqrt{\frac{s+4m^2_\text{E}}{s}} \ \theta(\sqrt{-s} -2 m_\text{E}) \  \left[ 
    D_{ \mathrm{F} }(p_2 - k_2) + D_{ \mathrm{F} }(p_2 - k_1) 
    \right]\left[
    D^\ast _{ \mathrm{F} }(\bar{p}_2 - k_2) + D^\ast _{ \mathrm{F} }(\bar{p}_2 - k_1)
    \right],
\end{align}
where $\sqrt{-s}=2\omega_{\bm{p}_1}$. 
Note that the four-momentum $k^\mu_2$ is given as $k^\mu_2 =(E_{k_1}, - \bm{k}_1)$ and the integral measure $d^3k_1$ can be rewritten by $E_{k_1}\sqrt{ E^2_{k_1} - m^2_\text{E} }dE_{k_1}d\Omega$.
Under the condition $|s| \ll m^2_\text{E}$, the step function $\theta(\sqrt{-s} - 2m_\text{E})$ is zero and the coefficient Eq.\eqref{eq:gamma_pair} vanishes.
Thus, the dissipation term $D_{\phi\phi \to \chi\chi}[\rho^\text{in}_\text{E}]$ is ignored, and the dynamics gets close to a unitary evolution.

We finally mention on the symmetry of GKSL generator. Through a straightforward way, we find that the generator has Poincar\'e symmetry in the sense that 
\begin{equation}
\hat{U} \mathcal{L}_{\phi\phi} [\rho^\text{in}_{\mathrm{s}}] \hat{U}^{\dagger} = \mathcal{L}_{\phi\phi} \left[ \hat{U} \rho^\text{in}_{\mathrm{s}}\hat{U}^{\dagger} \right],
\end{equation}
werer $\hat{U}=\hat{U}(\Lambda,a)$ is the unitary representation of Poincare transformation. 
This is checked in Appendix \ref{app:Confirm_symmetry} by using the Poincar\'{e} transformation rule of $\hat{\mathrm{a}}^\dagger_{\bm{p} \bm{p}'}$:
\begin{align}
    \hat{U} \hat{ \mathrm{a} }^{\dagger}_{\bm{p}\,\bm{p}'} \hat{U}^{\dagger}
    = \sqrt{ \frac{\omega_{\bm{p}_\Lambda}\omega_{\bm{p}'_\Lambda}}{\omega_{\bm{p}}\omega_{\bm{p}'}}}
    e^{ -i(\Lambda p+\Lambda p')^\mu a_\mu }
    \hat{ \mathrm{a} }^{\dagger}_{ \bm{p}_\Lambda \bm{p}'_\Lambda}.
    \label{eq:Poincare_trans2}
\end{align}
This means that the open dynamics of $\phi$ particles in the pair annihilation and the $2\to 2$ scattering processes is effectively described by the GKSL generator with the Poincar\'e symmetry.
\begin{table}[H]
    \centering
    \begin{tabular}{|c|}
    \hline
     The properties of $\mathcal{L}_{\phi\phi}[\rho^\text{in}_\text{s}]$ \\ \hline
     \begin{minipage}{0.8\textwidth}
         \vspace{5pt}
         \begin{enumerate}
             \item This generator has Poincar\'e symmetry.
             \item Two characteristic features:
             \begin{enumerate}
                 \item Interaction by the exchange of virtual particles
                 \item Particle number decrease by pair annihilation and its probability depending on the relative phase of incident superposed particles
             \end{enumerate}
             \item If $m_\text{s} \ll m_\text{E}$, the pair annihilation process is ignored and the dynamics gets unitary.
         \end{enumerate}
         \vspace{3pt}
     \end{minipage} \\ \hline
    \end{tabular}
    \caption{Properties of $\mathcal{L}_{\phi\phi}[\rho^\text{in}_\text{s}]$}
    \label{tab:property_pair}
\end{table}

\section{Discussion}
\label{sec:Discussion}
In Sec.\ref{sec:formulation_GKSL}, we derived the GKSL generators by focusing on the decay of particle, and the pair annihilation of particles and the $2\to 2$ scattering. 
These generators have the Markovian property since they are given by the GKSL form. 
Conventionally, the GKSL generator in the theory of OQS is derived under three approximations: the Born approximation, the Markov approximation and the rotating wave approximation\cite{Breuer2002}.  
However, in the derivation of Eq.\eqref{eq:L_scatt}, it seems not to use these approximations.
So, it is interesting to discuss why the GKSL generator appears in scattering process. 
Firstly, we consider the Born approximation. 
In Sec.\ref{sec:formulation_GKSL}, we expanded the S-operator as 
\begin{align}
    \hat{S} 
    = \hat{ \mathbb{I} } + i \hat{T}
    = \hat{ \mathbb{I} } + \sum^{\infty}_{n=1} 
    \frac{(-i)^{n}}{n!} \int^{\infty}_{-\infty} dt_1 \int^{\infty}_{-\infty} dt_2 \cdots \int^{\infty}_{-\infty} dt_n
    \mathrm{T} \left[ \hat{V}_{\mathrm{I}}(t_1)\hat{V}_{\mathrm{I}}(t_2) \cdots \hat{V}_{\mathrm{I}}(t_n) \right] \notag
\end{align}
The operator $\hat{T}$ includes the all effects of interaction, but we only took the leading order in exemplifying the GKSL generator.  
We think that this procedure corresponds to the Born approximation. 
Secondly, we discuss the Markov approximation. 
As observed in \eqref{eq:S} and \eqref{eq:S-dynamics}, we considered the OQS $\phi$ coupled to the environment $\chi$ and focused on the dynamics of $\phi$ from the infinite past  $t=-\infty$ to the infinite future $t=\infty$. 
To make the Markov approximation valid, the correlation time $\tau_{\mathrm{E}}$ of environment 
(the time that environment correlates with OQS) need to be much smaller than the evolving time $\tau_{\mathrm{s}}$ of OQS, that is, $\tau_{\mathrm{E}} \ll \tau_{\mathrm{s}}$. 
After the decay process or the pair annihilation process considered in this study, $\chi$ particles generated from $\phi$ particles fly away to spatial infinity. 
Hence, the state of $\chi$ correlates with the state of $\phi$ only during the typical time of each process. 
The Markov approximation would be guaranteed by focusing on the evolution of $\phi$ over a much longer time than such a typical time.
Finally, we consider the rotating wave approximation. This is the approximation that the terms rapidly oscillating are neglected. 
The rotating wave approximation would be valid because the oscillating terms are averaged to be zero by pursuing the (infinitely) long time behavior of $\phi$. 
These are the reason that the GKSL generator of $\phi$ particles is obtained in this study. 
Of course, the above discussion is still speculative, so it is necessary to compare the generators derived after imposing these approximations with those in this study to be rigorously sure.

Next, let us consider the pair annihilation probability derived in Sec.\ref{sec:Pair_annihilation}.
As can be seen Fig.\eqref{fig:Prob_0.01}, as the phase difference $\delta$ increases, the peak of probability goes down.
We discuss what is happening in considering the incident superposed particles $|\bm{q},-\bm{q} \rangle_\text{s} +e^{i\delta} |\bar{\bm{q}},-\bar{\bm{q}} \rangle_\text{s}$.
A pair of $\chi$ particles is produced through two distinct processes: the pair annihilation process of $\phi$ particles in the state $|\bm{q},-\bm{q}\rangle_\text{s}$ and in the state $|\bm{\bar{q}},-\bm{\bar{q}}\rangle_\text{s}$.
Since the pair of $\phi$ particles is superposed, the pair of $\chi$ particles produced in each process would be superposed and interfere with each other. 
Therefore, the pair annihilation probability reflects the interference effects and depends on the relative phase $\delta$. 
In our demonstration, $\delta = 0$ strengthens the pair annihilation process, while $\delta = \pi$ weakens it.
This would explains why the behavior observed in Fig.\ref{fig:Prob_0.01}.
\begin{figure}[H]
    \centering
    \includegraphics[width=0.3\linewidth]{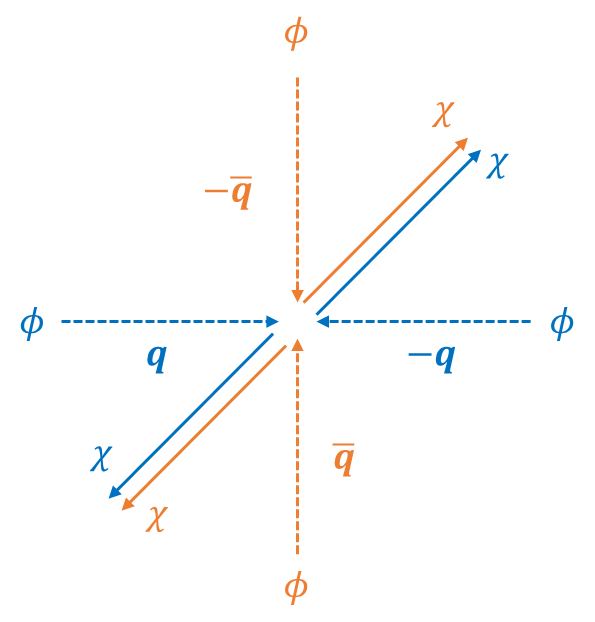}
    \caption{
    A schematic illustration of the pair annihilation of $\phi$ particles and the pair production of $\chi$ particles is shown.  
    The state $|\bm{q}, -\bm{q} \rangle_\text{s}$ of $\phi$ particles corresponds to horizontal incidence, represented by the pair of blue dotted arrows. Similarly, the state $|\bm{\bar{q}}, -\bm{\bar{q}} \rangle_\text{s}$ corresponds to vertical incidence, shown as the pair of orange dotted arrows.  
    From each incidence event, a pair of $\chi$ particles is produced. These events are quantum mechanically superposed, resulting in interference between the produced pairs of $\chi$ particles, i.e. between the pair of blue solid arrows and the pair of orange solid arrows.
    }
    \label{fig:Prob_discuss}
\end{figure}

Here, we compare the present results with those of our previous research \cite{Kaito_2024}. In Ref.\cite{Kaito_2024}, we derived the relativistic GKSL equation with Poincar\'e symmetry for spin 0 massive particle. The GKSL equation in the previous research is given as
\begin{align}
    \frac{d}{dt} \rho(t) = \mathcal{L}_\text{pre} [\rho(t)], \quad \mathcal{L}_\text{pre}[\rho(t)]
    = -i \left[ \hat{H} + g\hat{N}, \rho(t) \right]
    + \gamma \int d^{3}p \left[ \hat{a}(\bm{p})\rho(t)\hat{a}^{\dagger}(\bm{p}) - \frac{1}{2} \left\{ \hat{a}^{\dagger}(\bm{p})\hat{a}(\bm{p}), \rho(t) \right\} \right],
    \label{eq:L_pre}
\end{align}
where $\hat{a}(\bm{p})$ is the annihilation operator of massive particle, 
$\hat{H}$ and $\hat{N}$ are given as
\begin{align}
    \hat{H} = \int d^3 p \ \omega_{\bm{p}} \hat{a}^{\dagger}(\bm{p}) \hat{a}(\bm{p}), \ 
    \hat{N} = \int d^3 p \ \hat{a}^{\dagger}(\bm{p})\hat{a}(\bm{p}).
\end{align}
At this point, there was a question that this GKSL equation appears from what kind of a physical system-environment model.    
We think that this present study can give the answer for this question. 
The above GKSL equation describes the decay of massive particle, and hence the particle must decay for an infinitely long time, whose state evolves to a vacuum state $|0\rangle \langle 0|$. 
As observed in Sec.\ref{sec:Decay_process}, the GKSL generator $\mathcal{L}_\phi$ diverges, and it suggests that $\phi$ particle must decay for the infinite time and its state transits to the vacuum state $|0\rangle_\text{s} \langle 0|$.
This means that the theory of GKSL equation with Poincar\'e symmetry is effective for describing the long time dynamics of decaying scalar particle.
This is the answer for the devoted question.

\section{Conclusion and outlook}
\label{sec:conclusion and outlook}
In this study, we investigated the open dynamics of quantum particles in scattering process. 
Based on the scattering theory, we derived the general form of GKSL generator and obtained the examples of the generator by considering the decay of scalar particle ($\phi \rightarrow \chi\chi$), and the pair annihilation ($\phi\phi \rightarrow \chi\chi$) and the $2\rightarrow2$ scattering of scalar particles ($\phi\phi \rightarrow \phi\phi$).
For the decay process, this process is determined by a decay rate \eqref{eq:decay_rate}, and the GKSL generator in this process asymptotically has Poincar\'e symmetry and describes that the particle always decays after an infinite time.
For the pair annihilation and the $2\rightarrow 2$ scattering processes, the GKSL generator Eq.\ref{eq:L_pair} also respects Poincar\'e symmetry and reflects the two behaviors: one is the interaction by the exchange of virtual particles $\chi$, and the other is the transition to the vacuum state through pair annihilation.
In particular, our investigation on the pair annihilation reveals that its probability varies with the relative phase in the superposition state of incident particles.
Overall, it was shown that the description by GKSL generator with Poincar\'e symmetry is effective for the asymptotic open  dynamics of quantum particles considered in this study.


In this paper, we think that our research have proposed one of methods to understand relativistic phenomena from the theory of OQS. 
In addition, GKSL generator in the theory of OQS can give dynamical map or quantum channel, which play an important role for the transmission of quantum information. 
This fact could give the way to understand relativistic phenomena by the approaches of quantum information theory.  
That being said, it is necessary to clarify how insights into relativistic phenomena can be obtained from the theory of OQS and quantum information.


As we mentioned in Sec.\ref{sec:Intro}, our research could be applied for exploring
quantum gravity theory in weak gravity regime.
In our previous research\cite{Kaito_2024}, we proposed the theory of relativistic GKSL equation with Poincar\'e symmetry towards establishing the comprehensive approach for 
quantum gravity theory.
However, it is still unclear how to incorporate gravitational interactions into the theory.
Applying this study for the concrete process via gravitational interactions would solve that issue. 
We hope that the present study gives a novel route for revealing quantum gravity theory. 

\begin{acknowledgements}
We thank D. Carney, H. Furugori, K. Gallock-Yoshimura, S. Kanno, J. Oppenheim and K. Yamamoto for variable discussions and comments related to this paper. 
A.M. was supported by JSPS KAKENHI (Grants No.~JP23K13103 and No.~JP23H01175).
\end{acknowledgements}

\begin{appendix}

\section{Derivation of Eq.\eqref{eq:S-dynamics}}
\label{app:drv1}
Here, we derive Eq.\eqref{eq:S-dynamics2} by substituting Eq.\eqref{eq:T-operator} for Eq.\eqref{eq:S-dynamics}:
\begin{align}
    \rho^{\mathrm{out}}_{ \mathrm{s} } 
    &= \mathrm{Tr}_\text{E} \left[ ( \hat{\mathbb{I}} + i \hat{T} ) ( \rho^\text{in}_{\mathrm{s}} \otimes |0 \rangle _{\mathrm{E}} \langle 0| ) ( \hat{\mathbb{I}} - i \hat{T}^{\dagger} ) \right] \notag \\
    &= \mathrm{Tr}_\text{E} \left[ \rho^\text{in}_{\mathrm{s}} \otimes |0 \rangle _{\mathrm{E}} \langle 0| + i\hat{T}( \rho^\text{in}_{\mathrm{s}} \otimes |0 \rangle _{\mathrm{E}} \langle 0| )  \right. \notag \\
    & \qquad \qquad \qquad  \left. -i(\rho^\text{in}_{\mathrm{s}} \otimes |0 \rangle _{\mathrm{E}} \langle 0|)\hat{T}^{\dagger} + \hat{T}( \rho^\text{in}_{\mathrm{s}} \otimes |0 \rangle _{\mathrm{E}} \langle 0| ) \hat{T}^{\dagger} \right] \notag \\
    &= \rho^\text{in}_{\mathrm{s}} + i \hat{T}_{0} \rho^\text{in}_{\mathrm{s}} -i \rho^\text{in}_{\mathrm{s}} \hat{T}^{\dagger}_{0} + \int d\beta \ \hat{T}_{\beta} \rho^\text{in}_{\mathrm{s}} \hat{T}^{\dagger}_{\beta} \notag \\
    &= \rho^{\mathrm{in}}_{\mathrm{s}} + i\left[ \mathrm{Re}\hat{T}_0, \rho^{\mathrm{in}}_{\mathrm{s}} \right] 
    - \left\{ \mathrm{Im}\hat{T}_0, \rho^\text{in}_{\mathrm{s}} \right\}
    + \int d\beta \ \hat{T}_{\beta} \rho^\text{in}_{\mathrm{s}} \hat{T}^{\dagger}_{\beta}
\end{align}
where we used the definition Eq.\ref{eq:def_T},
\begin{align}
    \hat{T}_{\beta} = {}_{\text{E}}\langle \beta | \hat{T} |0 \rangle_\text{E}, \quad \hat{T}_{0} = {}_\text{E}\langle 0| \hat{T} | 0 \rangle_\text{E}.
\end{align}
Furthermore, we can rewrite this result by using the unitary condition of S-operator. The unitary condition can be written as
\begin{align}
    \hat{ \mathbb{I} } 
    = \hat{S}^\dagger\hat{S}
    = \hat{ \mathbb{I} } + i(\hat{T} - \hat{T}^\dagger) + \hat{T}^\dagger \hat{T},
    \hspace{1cm} \therefore \ i(\hat{T} - \hat{T}^\dagger) + \hat{T}^\dagger \hat{T} = 0
\end{align}
Sandwiching this equation by ${}_\text{E} \langle 0|$ and $|0\rangle_\text{E}$, and 
inserting the completeness condition $\hat{ \mathbb{I} }_{ \mathrm{E} } = \int d\beta |\beta\rangle_{\mathrm{E}}\langle \beta|$ between $\hat{T}^\dagger$ and $\hat{T}$ of the term $\hat{T}^\dagger \hat{T}$, we obtain
\begin{align}
    - 2 \mathrm{Im}\hat{T}_0 + \int d\beta \ \hat{T}^{\dagger}_\beta \hat{T}_\beta = 0, \quad \therefore \
    \mathrm{Im}\hat{T}_{0} = \frac{1}{2} \int d\beta \ \hat{T}^{\dagger}_{\beta} \hat{T}_{\beta}, 
\end{align}
where $\text{Im}[\hat{T}_0]=(\hat{T}_0 -\hat{T}^\dagger_0)/(2i)$. 
Therefore, Eq.\eqref{eq:S-dynamics} is rewritten as
\begin{align}
    \rho^{\mathrm{out}}_{\mathrm{s}}=
    \rho^{\mathrm{in}}_{\mathrm{s}} + i\left[ \mathrm{Re}\hat{T}_0, \rho^{\mathrm{in}}_{\mathrm{s}} \right] 
    + \int d\beta \ \left[ \hat{T}_{\beta} \rho^\text{in}_{\mathrm{s}} \hat{T}^{\dagger}_{\beta} - \frac{1}{2} \left\{ \hat{T}^{\dagger}_{\beta} \hat{T}_{\beta}, \rho^{\mathrm{in}}_{\mathrm{s}} \right\} \right].
\end{align}
Here, we only used the fact that the initial state of environment is vacuum and the S-operator is unitary.

\section{Confirmation of Poincar\'e symmetry}
\label{app:Confirm_symmetry}
In this section, we confirm that the GKSL generators derived in Sec.\ref{sec:Examples_GKSL} have Poincar\'e symmetry.
Here, we reprint the condition equation of Poincar\'e symmetry for GKSL generator.
\begin{align}
    \hat{U} \mathcal{L}_{\mathrm{scatt}}[\rho^\text{in}_{\mathrm{s}}] \hat{U}^{\dagger}
    = \mathcal{L}_{\mathrm{scatt}} \left[ \hat{U}\rho^\text{in}_{\mathrm{s}}\hat{U}^{\dagger} \right],
    \label{eq:Con_Pore}
\end{align}
where $\hat{U}=\hat{U}(\Lambda,a)$ is the unitary representation of Poincar\'e transformation. By substituting each GKSL generator for Eq.\eqref{eq:Con_Pore}, the Poincar\'e symmetry can be confirmed.
Firstly, we show the Poincar\'e symmetry of Eq.\eqref{eq:L_decay2},
\begin{align}
    \hat{U}\mathcal{L}_{\phi}[\rho^\text{in}_{\mathrm{s}}] \hat{U}^{\dagger}
    &=  \gamma 
     \int \frac{d^3 p d^3 \bar{p}}{\sqrt{ \omega_p \omega_{\bar{p}} }} \ \delta^4(p - \bar{p} )  \hat{U} \left[ 
   \hat{ \mathrm{a} }_{ \bm{p} }\rho^\text{in}_{\mathrm{s}}\hat{ \mathrm{a} }^{\dagger}_{\bm{ \bar{p} }} - \frac{1}{2} 
    \left\{ 
    \hat{ \mathrm{a} }^{\dagger}_{\bm{ \bar{p} } }\hat{ \mathrm{a} }_{ \bm{p} }
    , \rho^\text{in}_{\mathrm{s}}
    \right\} 
    \right] \hat{U}^\dagger\notag \\
    &= \gamma
    \int \frac{d^3 p d^3 \bar{p}}{\sqrt{ \omega_p \omega_{\bar{p}} }} \ \delta^4(p - \bar{p} ) 
    \sqrt{ \frac{ \omega_{\bm{p}_{\Lambda}} \omega_{\bar{\bm{p}}_{\Lambda}} }{\omega_{\bm{p}} \omega_{\bar{\bm{p}}}} } e^{-i(\Lambda p -\Lambda \bar{p})^\mu a_\mu}\left[ 
    \hat{ \mathrm{a} }_{ \bm{p}_\Lambda} \hat{U}\rho^\text{in}_{\mathrm{s}}\hat{U}^\dagger
    \hat{ \mathrm{a} }^{\dagger}_{\bm{ \bar{p} }_\Lambda}  
    - \frac{1}{2} 
    \left\{ 
    \hat{ \mathrm{a} }^{\dagger}_{\bm{ \bar{p}_\Lambda } }
    \hat{ \mathrm{a} }_{ \bm{p}_\Lambda }
    , \hat{U}\rho^\text{in}_{\mathrm{s}}\hat{U}^\dagger
    \right\} 
    \right] \notag \\
    &= \gamma
    \int \frac{d^3 p d^3 \bar{p}}{\sqrt{ \omega_{p} \omega_{\bar{p}}}} \ \delta^4(p -\bar{p} )  \left[ 
    \hat{ \mathrm{a} }_{ \bm{p} }
    \hat{U}\rho^\text{in}_{\mathrm{s}}\hat{U}^\dagger
    \hat{ \mathrm{a} }^{\dagger}_{\bm{ \bar{p} }}  
    - \frac{1}{2} 
    \left\{ 
    \hat{ \mathrm{a} }^{\dagger}_{\bm{ \bar{p}} }
    \hat{ \mathrm{a} }_{ \bm{p} }
    , \hat{U}\rho^\text{in}_{\mathrm{s}}\hat{U}^\dagger
    \right\} 
    \right] \notag \\
    &=\mathcal{L}_{\phi}\left[ \hat{U}\rho^\text{in}_{\mathrm{s}}\hat{U}^\dagger \right]
    \label{eq:Poi_decay},
\end{align}
where 
$\gamma=\frac{\lambda^2}{m_\text{s}} \sqrt{m_\text{s}^2 - 4m_\text{E}^2} \ \theta\left(m_\text{s}^2 - 4 m_\text{E}^2 \right)$. We used the transformation rule of $\hat{\mathrm{a}}_{\bm{p}}$ given in Eq.\eqref{eq:Poincare_trans} in the second line and the fact that $d^3p/\omega_{\bm{p}}$ is a Lorentz-invariant measure of integral in the third line.
Next, we check the Poincar\'e symmetry of Eq.\eqref{eq:L_pair}. By substituting Eq.\eqref{eq:L_pair} for Eq.\eqref{eq:Con_Pore}, we meet the following result:
\begin{align}
    &\hat{U} \mathcal{L}_{\phi\phi}[\rho^\text{in}_{\mathrm{s}}] \hat{U}^{\dagger} 
    \nonumber 
    \\
    &= \hat{U}\mathcal{H}[\rho^\text{in}_{\mathrm{s}}] \hat{U}^{\dagger}
    + \hat{U} \mathcal{D}[\rho^\text{in}_{\mathrm{s}}] \hat{U}^{\dagger} \notag \\
    &= i \frac{2\lambda^4}{(2\pi)^6} \int \frac{d^3 \bar{p} \, d^3 \bar{p}' d^3 p\, d^3 p'}{\sqrt{\omega_{\bar{\bm{p}}} \omega_{\bar{\bm{p}}'} \omega_{\bm{p}} \omega_{\bm{p}'}}}\mathrm{Im}\mathcal{A}(s,t,u) \delta^4 (\bar{p} + \bar{p}' - p - p ) 
     \hat{U}
     \left[ 
     \hat{\mathrm{a}}^{\dagger}_{\bm{\bar{p}}\bm{\bar{p}}'} 
     \hat{\mathrm{a}}_{\bm{p}\bm{p}'} ,
     \rho^\text{in}_{\mathrm{s}}
     \right]
     \hat{U}^\dagger \notag \\
     &\quad + \frac{4\lambda^4}{(2\pi)^4}  \int \frac{d^3 \bar{p} \, d^3 \bar{p}' d^3 p\, d^3 p'}{\sqrt{\omega_{\bar{\bm{p}}} \omega_{\bar{\bm{p}}'} \omega_{\bm{p}} \omega_{\bm{p}'}}} \ \gamma(p,p',\bar{p})\delta^4 (\bar{p} + \bar{p}' - p - p' ) 
    \hat{U}
    \left[ 
    \hat{\mathrm{a}}_{\bm{p}\bm{p}'} 
    \rho^\text{in}_{\mathrm{s}}
    \hat{\mathrm{a}}^{\dagger}_{\bm{\bar{p}}\bm{\bar{p}}'} 
    - \frac{1}{2} \left\{ 
    \hat{\mathrm{a}}^{\dagger}_{\bm{\bar{p}}\bm{\bar{p}}'} 
    \hat{\mathrm{a}}_{\bm{p}\bm{p}'} ,
    \rho^\text{in}_{\mathrm{s}}
    \right\}
    \right]
    \hat{U}^\dagger \notag \\
    &=i \frac{2\lambda^4}{(2\pi)^6} 
    \int \frac{d^3 \bar{p} \, d^3 \bar{p}' d^3 p\, d^3 p'}{\sqrt{\omega_{\bar{\bm{p}}} \omega_{\bar{\bm{p}}'} \omega_{\bm{p}} \omega_{\bm{p}'}}} 
    \
    \mathrm{Im}\mathcal{A}(s,t,u)
    \delta^4 (\bar{p} + \bar{p}' - p - p' ) \notag \\
    &\quad \times
     \sqrt{ \frac{ \omega_{\bar{\bm{p}}_{\Lambda}} \omega_{\bar{\bm{p}}'_{\Lambda}}\omega_{\bm{p}_{\Lambda}} \omega_{\bm{p}'_{\Lambda}}}{\omega_{\bar{\bm{p}}} \omega_{\bar{\bm{p}}'}\omega_{\bm{p}} \omega_{\bm{p}'_{\Lambda}}}} 
     e^{-i(\Lambda \bar{p} +\Lambda \bar{p}'-\Lambda p -\Lambda p')^\mu a_\mu}
     \left[ \hat{\mathrm{a}}^{\dagger}_{\bm{\bar{p}}_\Lambda \bm{\bar{p}}'_\Lambda}
     \hat{\mathrm{a}}_{\bm{p}_\Lambda \bm{p}'_\Lambda }, \hat{U}\rho^\text{in}_{\mathrm{s}}\hat{U}^\dagger
     \right] \notag \\
    &\quad + \frac{4\lambda^4}{(2\pi)^4}   \int \frac{d^3 \bar{p} \, d^3 \bar{p}' d^3 p\, d^3 p'}{\sqrt{\omega_{\bar{\bm{p}}} \omega_{\bar{\bm{p}}'} \omega_{\bm{p}} \omega_{\bm{p}'}}}  
    \ 
    \gamma(p,p',\bar{p})\delta^4 (\bar{p} + \bar{p}' - p - p' ) \notag \\
    &\quad
    \times 
     \sqrt{ \frac{ \omega_{\bar{\bm{p}}_{\Lambda}} \omega_{\bar{\bm{p}}'_{\Lambda}}\omega_{\bm{p}_{\Lambda}} \omega_{\bm{p}'_{\Lambda}}}{\omega_{\bar{\bm{p}}} \omega_{\bar{\bm{p}}'}\omega_{\bm{p}} \omega_{\bm{p}'_{\Lambda}}}} 
     e^{-i(\Lambda \bar{p} +\Lambda \bar{p}'-\Lambda p -\Lambda p')^\mu a_\mu}
    \left[ 
    \hat{\mathrm{a}}_{\bm{p}_\Lambda \bm{p}'_\Lambda} \hat{U}\rho^\text{in}_{\mathrm{s}}\hat{U}^\dagger \hat{\mathrm{a}}^{\dagger}_{\bm{\bar{p}}_\Lambda \bm{\bar{p}}'_\Lambda}
    - \frac{1}{2} \left\{ \hat{\mathrm{a}}^{\dagger}_{\bm{\bar{p}}_\Lambda \bm{\bar{p}}'_\Lambda}
    \hat{\mathrm{a}}_{\bm{p}_\Lambda \bm{p}'_\Lambda}, \hat{U}\rho^\text{in}_{\mathrm{s}}\hat{U}^\dagger
    \right\}
    \right] \notag \\
    &=i \frac{2\lambda^4}{(2\pi)^6} 
    \int \frac{d^3 \bar{p} \, d^3 \bar{p}' d^3 p\, d^3 p'}{\sqrt{\omega_{\bar{\bm{p}}} \omega_{\bar{\bm{p}}'} \omega_{\bm{p}} \omega_{\bm{p}'}}} 
    \
    \mathrm{Im}\mathcal{A}(s,t,u)
    \delta^4 (\bar{p} + \bar{p}' - p - p' )
     \left[ \hat{\mathrm{a}}^{\dagger}_{\bm{\bar{p}}\bm{\bar{p}}'}
     \hat{\mathrm{a}}_{\bm{p} \bm{p}'}, \hat{U}\rho^\text{in}_{\mathrm{s}}\hat{U}^\dagger
     \right] \notag \\
    &\quad + \frac{4\lambda^4}{(2\pi)^4}   \int \frac{d^3 \bar{p} \, d^3 \bar{p}' d^3 p\, d^3 p'}{\sqrt{\omega_{\bar{\bm{p}}} \omega_{\bar{\bm{p}}'} \omega_{\bm{p}} \omega_{\bm{p}'}}}  
    \ 
    \gamma(p,p',\bar{p})\delta^4 (\bar{p} + \bar{p}' - p - p' )
    \left[ 
    \hat{\mathrm{a}}_{\bm{p}\bm{p}'} \hat{U}\rho^\text{in}_{\mathrm{s}}\hat{U}^\dagger \hat{\mathrm{a}}^{\dagger}_{\bm{\bar{p}} \bm{\bar{p}}'}
    - \frac{1}{2} \left\{ \hat{\mathrm{a}}^{\dagger}_{\bm{\bar{p}} \bm{\bar{p}}'}
    \hat{\mathrm{a}}_{\bm{p} \bm{p}'}, \hat{U}\rho^\text{in}_{\mathrm{s}}\hat{U}^\dagger
    \right\}
    \right] \notag \\
    &= \mathcal{H}[\hat{U}\rho^\text{in}_{\mathrm{s}}\hat{U}^\dagger] 
    + \mathcal{D}[\hat{U}\rho^\text{in}_{\mathrm{s}}\hat{U}^\dagger]
     \notag \\
    &= \mathcal{L}_{\phi\phi}[\hat{U}\rho^\text{in}_{\mathrm{s}}\hat{U}^\dagger],
    \label{eq:Poi_pair}
\end{align}
where in the third equality the transformation rule of $\hat{\mathrm{a}}_{\bm{p}\bm{p}'}$ in Eq.\eqref{eq:Poincare_trans} was used, and in the fourth equality note that $\mathcal{A}(s,t,u)$ and $\gamma(p,p',\bar{p})$ are invariant under the Lorentz transformation $p \to \Lambda p, p'\to \Lambda p', \bar{p} \to \Lambda \bar{p}$ and $\bar{p}' \to \Lambda \bar{p}'$. 

\section{Mass renormalization}
\label{app:renormalization}
In Sec.\ref{sec:Examples_GKSL}, we mentioned the diagrams removed by renormalization. In this section, we explain the renormalization used in Sec.\ref{sec:Decay_process} as an example. Introducing the counter term of mass proportional to $\delta m^2_\text{s}$, we have the total Hamiltonian $\hat{H}_{\mathrm{tot}}$,
\begin{align}
    \hat{H}_{ \mathrm{tot} } 
    = \hat{H}_{ \mathrm{s} } \otimes \hat{ \mathbb{I} }_{ \mathrm{E} } +
    \hat{\mathbb{I}}_{ \mathrm{s} } \otimes \hat{H}_{ \mathrm{E} }
    + \hat{V}_{ \mathrm{I} }
    + \delta m_\text{s}^2 \int d^3 x \ \hat{\phi}^2 \otimes \hat{\mathbb{I}}_{\mathrm{E}} 
    = \hat{H}^0_{\mathrm{tot}} + \hat{V}'_{\mathrm{I}},
\end{align}
where the new operators $\hat{H}^0_{\mathrm{tot}}$, $\hat{V}'_{\mathrm{I}}$ are given as 
\begin{align}
    \hat{H}^0_{\mathrm{tot}} &= \hat{H}_{ \mathrm{s} } \otimes \hat{ \mathbb{I} }_{ \mathrm{E} } +
    \hat{\mathbb{I}}_{ \mathrm{s} } \otimes \hat{H}_{ \mathrm{E} },
    \label{eq:free_H}
    \\
    \hat{V}'_{\mathrm{I}} &= \hat{V}_{\mathrm{I}} + 
    \delta m_\text{s}^2 \int d^3 x \ \hat{\phi}^2 \otimes \hat{\mathbb{I}}_{\mathrm{E}}
    \label{eq:new_VI}
\end{align}
Hereafter, it does not matter if where we take the origin of energy, and we take the normal ordering of field $\phi$. 
According to the Feynmann diagram Fig.\ref{fig:one-loop}, the amplitude ${}_{\mathrm{E}}\langle 0 | {}_{\mathrm{s}}\langle \bar{\bm{p}} | \ i\hat{T} \ | \bm{p} \rangle_{\mathrm{s}} |0 \rangle_{\mathrm{E}}$ without the counter term is 
\begin{align}
   {}_{\mathrm{E}}\langle 0 | {}_{\mathrm{s}}\langle \bar{\bm{p}} | \ i\hat{T} \ | \bm{p} \rangle_{\mathrm{s}} |0 \rangle_{\mathrm{E}}
   &= \frac{(2\pi)^{ -\frac{3}{2} }}{ \sqrt{2\omega_p} } \cdot \frac{(2\pi)^{ -\frac{3}{2} } }{ \sqrt{2\omega_{\bar{p}}} } \cdot
   \int d^4 q \int d^4 \bar{q} \ 
   (-i\lambda) (2\pi)^4 \delta^4(q + \bar{q} - p) \cdot (-i\lambda) (2\pi)^4 \delta^4(q + \bar{q} - \bar{p}) \notag \\
   &\quad \times \frac{-i}{(2\pi)^4} \frac{1}{q^2 + m_\text{E}^2 - i\epsilon} \cdot \frac{-i}{(2\pi)^4} \frac{1}{\bar{q}^2 + m_\text{E}^2 - i\epsilon} \notag \\
   &= \frac{\lambda^2 \mathcal{B}}{ \sqrt{\omega_p \omega_{\bar{p}} } } \ \delta^4(p - \bar{p}) , 
\end{align}
where $\mathcal{B}$ is 
\begin{align}
  \mathcal{B}
   = \frac{1}{2(2\pi)^3} \int  \frac{d^4 q}{[q^2 + m_\text{E}^2 - i\epsilon][(q-p)^2 + m_\text{E}^2 - i\epsilon]}. 
\end{align}
This $d^4q$ integral leads to the logarithmic UV divergence. Note that the quantity $\mathcal{B}$ is constant independent of the momentum $\bm{p}$ since it is Lorentz invariant under $p^\mu \to (\Lambda p)^\mu$.  
In the following $\mathcal{B}$ is assumed to be regularized by introducing a UV cutoff parameter. For example, we adopt the dimensional regularization consistent with the Poincar\'e symmetry.   
We get the operator $\hat{T}_0$
\begin{align}
   \hat{T}_0
   = -i \lambda^2 \mathcal{B}\int \frac{d^3pd^3\bar{p}}{\sqrt{\omega_{\bm{p}} \omega_{\bar{\bm{p}}}}} \ \delta^4(p-\bar{p})
    \hat{ \mathrm{a} }^\dagger_{ \bm{\bar{p}} } \hat{ \mathrm{a} }_{ \bm{p} },
\end{align}
and its real part is important,
\begin{align}
    \mathrm{Re} \hat{T}_0 
    = \lambda^2 \text{Im}[\mathcal{B}]\int \frac{d^3pd^3\bar{p}}{\sqrt{\omega_{\bm{p}}\omega_{\bar{\bm{p}}}} } \ \delta^4(p-\bar{p})
    \hat{ \mathrm{a} }^\dagger_{ \bm{\bar{p}} } \hat{ \mathrm{a} }_{ \bm{p} }. 
\end{align}
On the other hand, the counter term given by the second term in Eq.\ref{eq:new_VI} gives
\begin{align}
    \delta \hat{T}_0  
    =  \delta m_\text{s}^2 \int d^4x \ \mathcal{N} \ [\hat{\phi}^2(x)] = 2\pi \delta m_\text{s}^2  \int \frac{d^3p d^3\bar{p}}{\sqrt{\omega_{\bm{p}} \omega_{\bar{\bm{p}}}} } \ \delta^4(p-\bar{p}) \
    \hat{ \mathrm{a} }^\dagger_{ \bm{\bar{p}} } \hat{ \mathrm{a} }_{\bm{p}}.
\end{align}
Note that the symbol $\mathcal{N}[\cdots]$ means the normal ordering. By choosing the parameter $\delta m_\text{s}^2$ as,
\begin{align}
    \lambda^2 \text{Im}\mathcal{B} + 2\pi \delta m_\text{s}^2 = 0,
\end{align}
the operator $\text{Re}\hat{T}_0$ can be set to zero. Therefore, the diagrams such as Fig.\ref{fig:one-loop} is removed by the mass renormalization. 
Even in the case of pair annihilation process, the diagrams with one-loop diagrams are removed by following the same discussion as above.

\section{Redefinition of S-operator}
\label{app:Re_Soperator}
In this section, we explain that vacuum bubble diagrams are removed. Because the vacuum state $|0\rangle_\phi \otimes |0\rangle_\chi$ is Lorentz invariant, the following relation holds:
\begin{align}
    \hat{S} |0\rangle_\phi \otimes |0\rangle_\chi
    = e^{i\theta} |0\rangle_\phi \otimes |0\rangle_\chi,
\end{align}
where $\hat{S}$ is S-operator and the parameter $\theta$ is a real phase. By taking the inner product with ${}_\phi\langle 0| \otimes {}_\chi\langle 0|$, the phase is obtained as
\begin{align}
    e^{i\theta} = {}_\phi\langle 0| \otimes {}_\chi\langle 0|\hat{S}|0\rangle_\phi \otimes |0\rangle_\chi,
\end{align}
which means that 
the phase comes from the vacuum bubble diagrams. Redefining the S-operator as $e^{-i\theta}\hat{S}$, we can cancel out vacuum bubble diagrams. In fact, for the decay process of interest, the amplitude ${}_\phi\langle \bm{\bar{p}}| \otimes {}_\chi\langle 0|\hat{S}|\bm{p}\rangle_\phi \otimes |0\rangle_\chi$ is calculated as
\begin{align}
    {}_\phi\langle \bm{\bar{p}}| \otimes {}_\chi\langle 0|\hat{S}|\bm{p}\rangle_\phi \otimes |0\rangle_\chi 
    &= \ 
    \begin{tikzpicture}[baseline=-0.5ex, scale=0.25]
    \definecolor{darkgreen}{RGB}{0,100,0}
    \begin{feynman}
      \vertex (a) at (0,-2) ;
      \vertex (b) at (0,2) ;
      \vertex (i1) at (-1,0);
      \vertex (o) at (0,0);
      \diagram* {
        (a) -- [scalar] (b),
      };
    \end{feynman}
   \end{tikzpicture} 
   \ \times \left( 1 + \frac{1}{2} \
   \begin{tikzpicture}[baseline=-0.5ex, scale=0.25]
    \definecolor{darkgreen}{RGB}{0,100,0}
    \begin{feynman}
      \vertex (a) at (0,2) ;
      \vertex (b) at (0,-2) ;
      \vertex (i1) at (0,1);
      \vertex (i2) at (0.5,1.5);
      \vertex (i3) at (-0.5,1.5);
      \vertex (j1) at (0,-1);
      \vertex (j2) at (0.5,-1.5);
      \vertex (j3) at (-0.5,-1.5);
      \diagram* {
        (a) -- [quarter left] (i2) -- [quarter left] (i1) -- [quarter left] (i3) -- [quarter left] (a),
        (i1) -- [scalar] (j1),
        (b) -- [quarter right] (j2) -- [quarter right] (j1) -- [quarter right] (j3) -- [quarter right] (b),
      };
    \end{feynman}
   \end{tikzpicture}
   + 
   \begin{tikzpicture}[baseline=-0.5ex, scale=0.25]
    \definecolor{darkgreen}{RGB}{0,100,0}
    \begin{feynman}
      \vertex (a) at (0,1) ;
      \vertex (b) at (0,-1) ;
      \vertex (c) at (1,0) ;
      \vertex (d) at (-1,0) ;
      \diagram* {
        (a) -- [quarter left] (c) -- [quarter left] (b) -- [quarter left] (d) -- [quarter left] (a),
        (d) -- [scalar] (c)
      };
    \end{feynman}
   \end{tikzpicture}
   + \mathcal{O}(\lambda^4) 
   \right)
   + 
   \begin{tikzpicture}[baseline=-0.5ex, scale=0.25]
    \definecolor{darkgreen}{RGB}{0,100,0}
    \begin{feynman}
      \vertex (a) at (0,2) ;
      \vertex (b) at (0,-2) ;
      \vertex (c) at (0.75,0) ;
      \vertex (d) at (-0.75,0) ;
      \vertex (i1) at (0,0.75) ;
      \vertex (i2) at (0,-0.75) ;
      \diagram* {
        (a) -- [scalar] (i1) -- [quarter left] (c) -- [quarter left] (i2) -- [quarter left] (d) -- [quarter left] (i1),
        (i2) -- [scalar] (b),
      };
    \end{feynman}
   \end{tikzpicture}
   \times \left( 1 + \frac{1}{2} \
   \begin{tikzpicture}[baseline=-0.5ex, scale=0.25]
    \definecolor{darkgreen}{RGB}{0,100,0}
    \begin{feynman}
      \vertex (a) at (0,2) ;
      \vertex (b) at (0,-2) ;
      \vertex (i1) at (0,1);
      \vertex (i2) at (0.5,1.5);
      \vertex (i3) at (-0.5,1.5);
      \vertex (j1) at (0,-1);
      \vertex (j2) at (0.5,-1.5);
      \vertex (j3) at (-0.5,-1.5);
      \diagram* {
        (a) -- [quarter left] (i2) -- [quarter left] (i1) -- [quarter left] (i3) -- [quarter left] (a),
        (i1) -- [scalar] (j1),
        (b) -- [quarter right] (j2) -- [quarter right] (j1) -- [quarter right] (j3) -- [quarter right] (b),
      };
    \end{feynman}
   \end{tikzpicture}
   +
   \begin{tikzpicture}[baseline=-0.5ex, scale=0.25]
    \definecolor{darkgreen}{RGB}{0,100,0}
    \begin{feynman}
      \vertex (a) at (0,1) ;
      \vertex (b) at (0,-1) ;
      \vertex (c) at (1,0) ;
      \vertex (d) at (-1,0) ;
      \diagram* {
        (a) -- [quarter left] (c) -- [quarter left] (b) -- [quarter left] (d) -- [quarter left] (a),
        (d) -- [scalar] (c)
      };
    \end{feynman}
   \end{tikzpicture}
   + \mathcal{O}(\lambda^4) 
   \right) + (\mathrm{high \ order \ of \ \lambda}) \notag \\
   &= \underset{\ diagrams \ without \ vacuum \ bubbles}{
   \left( \
   \begin{tikzpicture}[baseline=-0.5ex, scale=0.25]
    \definecolor{darkgreen}{RGB}{0,100,0}
    \begin{feynman}
      \vertex (a) at (0,-2) ;
      \vertex (b) at (0,2) ;
      \vertex (i1) at (-1,0);
      \vertex (o) at (0,0);
      \diagram* {
        (a) -- [scalar] (b),
      };
    \end{feynman}
   \end{tikzpicture}
   \ + 
   \begin{tikzpicture}[baseline=-0.5ex, scale=0.25]
    \definecolor{darkgreen}{RGB}{0,100,0}
    \begin{feynman}
      \vertex (a) at (0,2) ;
      \vertex (b) at (0,-2) ;
      \vertex (c) at (0.75,0) ;
      \vertex (d) at (-0.75,0) ;
      \vertex (i1) at (0,0.75) ;
      \vertex (i2) at (0,-0.75) ;
      \diagram* {
        (a) -- [scalar] (i1) -- [quarter left] (c) -- [quarter left] (i2) -- [quarter left] (d) -- [quarter left] (i1),
        (i2) -- [scalar] (b),
      };
    \end{feynman}
   \end{tikzpicture}
   + \mathcal{O}(\lambda^4)
   \right)
   }
   \times \qquad
   \underset{ vacuum \ bubbles}{
   \left( 1 + \frac{1}{2} \
   \begin{tikzpicture}[baseline=-0.5ex, scale=0.25]
    \definecolor{darkgreen}{RGB}{0,100,0}
    \begin{feynman}
      \vertex (a) at (0,2) ;
      \vertex (b) at (0,-2) ;
      \vertex (i1) at (0,1);
      \vertex (i2) at (0.5,1.5);
      \vertex (i3) at (-0.5,1.5);
      \vertex (j1) at (0,-1);
      \vertex (j2) at (0.5,-1.5);
      \vertex (j3) at (-0.5,-1.5);
      \diagram* {
        (a) -- [quarter left] (i2) -- [quarter left] (i1) -- [quarter left] (i3) -- [quarter left] (a),
        (i1) -- [scalar] (j1),
        (b) -- [quarter right] (j2) -- [quarter right] (j1) -- [quarter right] (j3) -- [quarter right] (b),
      };
    \end{feynman}
   \end{tikzpicture}
   + 
   \begin{tikzpicture}[baseline=-0.5ex, scale=0.25]
    \definecolor{darkgreen}{RGB}{0,100,0}
    \begin{feynman}
      \vertex (a) at (0,1) ;
      \vertex (b) at (0,-1) ;
      \vertex (c) at (1,0) ;
      \vertex (d) at (-1,0) ;
      \diagram* {
        (a) -- [quarter left] (c) -- [quarter left] (b) -- [quarter left] (d) -- [quarter left] (a),
        (d) -- [scalar] (c)
      };
    \end{feynman}
   \end{tikzpicture}
   + \mathcal{O}(\lambda^4) 
   \right)}.
\end{align}
Thus the terms from vacuum bubbles are absorbed with the coefficient $e^{-i\theta}$, and the effects of vacuum bubbles are ignored by redefining the S-operator. 

\section{Diagrams removed for considering $2\to2$ scattering process}
\label{app:explanation of ReT_0 in 2-2}
In Sec.\ref{sec:Pair_annihilation}, we mentioned that the diagrams which give the amplitude $ {}_{\mathrm{E}}\langle 0 | {}_{\mathrm{s}}\langle \bm{\bar{p}}_1, \bm{\bar{p}}_2 | \ i \hat{T} \ | \bm{p}_1,\bm{p}_2 \rangle_{\mathrm{s}} | 0 \rangle _{\mathrm{E}}$ are only the diagrams shown in Fig.\ref{fig:4th-order diagram}. In this section, we explain this statement. 
Firstly, we focus on the diagrams with $O(\lambda^2)$. The scattering amplitude at this order consists of a kind of diagrams shown in Figs.\ref{fig:2nd-diagram with vac.} and \ref{fig:2nd-diagram without vac.}. 
\begin{figure}[H]
 \centering
  \begin{minipage}[b]{0.35\textwidth}
    \begin{tikzpicture}
    \centering
    \definecolor{darkgreen}{RGB}{0,100,0}
    \begin{feynman}
      \vertex (a) at (-2,-2) {$\phi$};
      \vertex (b) at (-2,2) {$\phi$};
      \vertex (c) at (-0.5,-2) {$\phi$};
      \vertex (d) at (-0.5,2) {$\phi$};
      \vertex (i1) at (-1.5,0.5) ;
      \vertex (i2) at (-1.5,-0.5) ;
      \vertex (i3) at (1,0) ;
      \vertex (i4) at (0,0) ;
      \vertex (j1) at (1,0.5) ;
      \vertex (j2) at (1,-0.5) ;
      \vertex (j3) at (0.5,0) ;
      \vertex (j4) at (1.5,0) ;
      \vertex (k1) at (-2,0.5);
      \vertex (k2) at (-2,-0.5);
      \vertex (l1) at (-0.5,0.5);
      \vertex (l2) at (-0.5,-0.5);
      \diagram* {
        (a) -- [scalar,momentum=\(p_1\)] (k2) -- [scalar] (k1) -- [scalar, momentum=\(\bar{p}_1\)] (b),
        (c) -- [scalar,momentum=\(p_2\)] (l2) -- [scalar] (l1) -- [scalar,momentum=\(\bar{p}_2\)] (d),
        (j1) -- [quarter left] (j4) -- [quarter left] (j2) -- [quarter left] (j3) -- [quarter left] (j1),
        (j1) -- [scalar] (j2),
      };
      \fill (j1) circle (2pt);
      \fill (j2) circle (2pt);
    \end{feynman}
  \end{tikzpicture}
  \end{minipage}
  \begin{minipage}[b]{0.35\textwidth}
    \centering
    \begin{tikzpicture}
    \definecolor{darkgreen}{RGB}{0,100,0}
    \begin{feynman}
      \vertex (a) at (-1,-2) {$\phi$};
      \vertex (b) at (-1,2) {$\phi$};
      \vertex (c) at (0.5,-2) {$\phi$};
      \vertex (d) at (0.5,2) {$\phi$};
      \vertex (i1) at (0.5,0.5) ;
      \vertex (i2) at (0.5,-0.5) ;
      \vertex (i3) at (1,0) ;
      \vertex (i4) at (0,0) ;
      \vertex (j1) at (2,1.5) ;
      \vertex (j2) at (2,0.5) ;
      \vertex (j3) at (1.5,1) ;
      \vertex (j4) at (2.5,1) ;
      \vertex (m1) at (2,-1.5) ;
      \vertex (m2) at (2,-0.5) ;
      \vertex (m3) at (1.5,-1) ;
      \vertex (m4) at (2.5,-1) ;
      \vertex (k1) at (-1,0.5);
      \vertex (k2) at (-1,-0.5);
      \vertex (l1) at (0.5,0.5);
      \vertex (l2) at (0.5,-0.5);
      \diagram* {
        (a) -- [scalar,momentum=\(p_1\)] (k2) -- [scalar] (k1) -- [scalar, momentum=\(\bar{p}_1\)] (b),
        (c) -- [scalar,momentum=\(p_2\)] (l2) -- [scalar] (l1) -- [scalar,momentum=\(\bar{p}_2\)] (d),
        (j1) -- [quarter left] (j4) -- [quarter left] (j2) -- [quarter left] (j3) -- [quarter left] (j1),
        (m1) -- [quarter right] (m4) -- [quarter right] (m2) -- [quarter right] (m3) -- [quarter right] (m1),
        (m2) -- [scalar] (j2),
      };
      \fill (j2) circle (2pt);
      \fill (m2) circle (2pt);
    \end{feynman}
  \end{tikzpicture}
  \end{minipage}
  \caption{Examples of the 2nd-order diagrams with vacuum bubbles in this process. These are absorbed by redefining the S-operator.}
  \label{fig:2nd-diagram with vac.}
\end{figure}
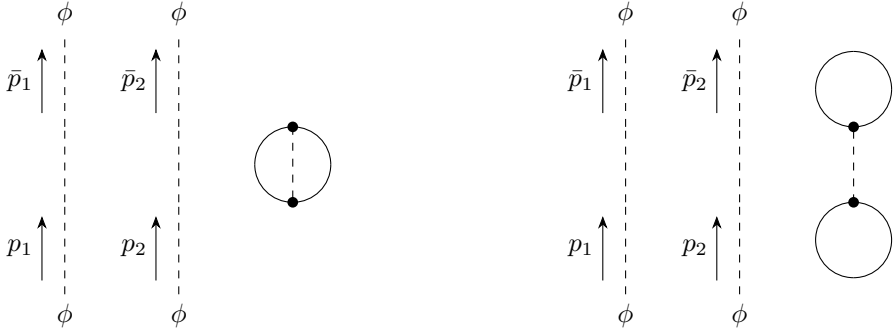
\begin{figure}[H]
\centering
  \begin{minipage}[b]{0.4\textwidth}
  \centering
    \begin{tikzpicture} 
    \definecolor{darkgreen}{RGB}{0,100,0}
    \begin{feynman}
      \vertex (a) at (-1,-2) {$\phi$};
      \vertex (b) at (-1,2) {$\phi$};
      \vertex (j1) at (-1,0.5);
      \vertex (j2) at (-1,-0.5);
      \vertex (c) at (1,-2) {$\phi$};
      \vertex (d) at (1,2) {$\phi$};
      \vertex (i1) at (1,-0.5);
      \vertex (i2) at (1,0.5);
      \vertex (i3) at (1.5,0) [label=right:{$\chi$}];
      \vertex (i4) at (0.5,0) [label=left:{$\chi$}];
      \diagram* {
        (a) -- [scalar, momentum=\(p_1\)] (j2) -- [scalar] (j1) -- [scalar, momentum=\(\bar{p}_1\)] (b),
        (c) -- [scalar,momentum={[shift={(-0.2, -0.2)}] \(p_2\)}] (i1) -- [quarter left] (i4) -- [quarter left] (i2) -- [quarter left] (i3) -- [quarter left] (i1),
        (i2) -- [scalar,momentum={[shift={(-0.2,0.2)}] \(\bar{p}_2\)}] (d),
      };
      \fill (i1) circle (2pt);
      \fill (i2) circle (2pt);
    \end{feynman}
  \end{tikzpicture}
  \end{minipage}
  \begin{minipage}[b]{0.4\textwidth}
    \begin{tikzpicture}
    \centering
    \definecolor{darkgreen}{RGB}{0,100,0}
    \begin{feynman}
      \vertex (a) at (-2,-2) {$\phi$};
      \vertex (b) at (-2,2) {$\phi$};
      \vertex (c) at (2,-2) {$\phi$};
      \vertex (d) at (2,2) {$\phi$};
      \vertex (i1) at (0.5,-0.5);
      \vertex (i2) at (-0.5,0.5);
      \vertex (i3) at (0.5,0.5) ;
      \vertex (i4) at (-0.5,-0.5) ;
      \vertex (j1) at (-1,-1); 
      \vertex (j2) at (1,1);
      \vertex (k1) at (-1,1);
      \vertex (k2) at (1,-1);
      \vertex (l1) at (-0.8,0) [label=left:{$\chi$}];
      \vertex (l2) at (0.8,0) [label=right:{$\chi$}];
      \diagram* {
        (a) -- [scalar, momentum=\(p_1\)] (j1) -- [scalar] (i4),
        (c) -- [scalar, momentum={[arrow distance=-3.0mm, label distance=-15pt] \(p_2\)}] (k2) -- [scalar] (k1) -- [scalar, momentum=\(\bar{p}_1\)] (b), 
        (i3) -- [scalar] (j2) -- [scalar,momentum={[arrow distance=-3.0mm, label distance=-15pt] \(\bar{p}_2\)}] (d),
        (i1) -- [quarter left] (i4) -- [quarter left] (i2) -- [quarter left] (i3) -- [quarter left] (i1),
      };
      \fill (i3) circle (2pt);
      \fill (i4) circle (2pt);
    \end{feynman}
  \end{tikzpicture}
  \end{minipage}
  \caption{Examples of the 2nd-order diagrams without vacuum bubbles in this process. These are removed by the mass renormalization.}
  \label{fig:2nd-diagram without vac.}
\end{figure}
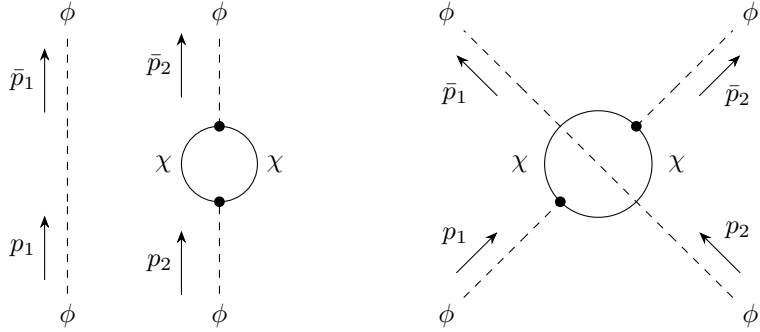

The diagrams with vacuum bubbles in Fig.\ref{fig:2nd-diagram with vac.} is absorved by the redefinition of S-operator as discussed in Appendix \ref{app:Re_Soperator}. 
The type of the diagrams in Fig.\ref{fig:2nd-diagram with vac.} includes a one-loop diagram is renormalized by the mass of OQS according to Appendix \ref{app:renormalization}, so they are ignored and ${}_{\mathrm{E}}\langle 0 | {}_{\mathrm{s}}\langle \bm{\bar{p}}_1, \bm{\bar{p}}_2 | \ i \hat{T} \ | \bm{p}_1,\bm{p}_2 \rangle_{\mathrm{s}} | 0 \rangle _{\mathrm{E}} $ at the second order vanishes.

Next, we will consider the fourth-order diagrams. Besides the diagrams as shown in Fig.\ref{fig:4th-order diagram}, we can consider some examples of diagrams depicted in Fig.\ref{fig:removed_diagram}.
\begin{figure}[H]
  \begin{minipage}[t]{0.33\textwidth}
    \centering
    \begin{tikzpicture}
    \definecolor{darkgreen}{RGB}{0,100,0}
    \begin{feynman}
      \vertex (a) at (-1,-2) {$\phi$};
      \vertex (b) at (-1,2) {$\phi$};
      \vertex (c) at (0.5,-2) {$\phi$};
      \vertex (d) at (0.5,2) {$\phi$};
      \vertex (i1) at (0.5,0.5) ;
      \vertex (i2) at (0.5,-0.5) ;
      \vertex (i3) at (1,0) ;
      \vertex (i4) at (0,0) ;
      \vertex (j1) at (2,0.5) ;
      \vertex (j2) at (2,-0.5) ;
      \vertex (j3) at (1.5,0) ;
      \vertex (j4) at (2.5,0) ;
      \vertex (k1) at (-1,0.5);
      \vertex (k2) at (-1,-0.5);
      \vertex (l1) at (0.5,0.5);
      \vertex (l2) at (0.5,-0.5);
      \diagram* {
        (a) -- [scalar,momentum=\(p_1\)] (k2) -- [scalar] (k1) -- [scalar, momentum=\(\bar{p}_1\)] (b),
        (c) -- [scalar,momentum=\(p_2\)] (l2) -- [scalar] (l1) -- [scalar,momentum=\(\bar{p}_2\)] (d),
        (j1) -- [quarter left] (j4) -- [quarter left] (j2) -- [quarter left] (j3) -- [quarter left] (j1),
        (j1) -- [scalar] (j2),
        (j3) -- [scalar] (j4),
      };
      \fill (j1) circle (2pt);
      \fill (j2) circle (2pt);
      \fill (j3) circle (2pt);
      \fill (j4) circle (2pt);
      \node[circle,inner sep=2pt] (A) at (2.8,2) {};
      \node[circle,inner sep=2pt] (B) at (2.8,-2) {};
    \end{feynman}
  \end{tikzpicture}
  \caption*{Redefinition of S-operator}
  \end{minipage}
   \begin{minipage}[t]{0.33\textwidth}
    \centering
    \begin{tikzpicture}
    \definecolor{darkgreen}{RGB}{0,100,0}
    \begin{feynman}
      \vertex (a) at (-0.5,-2) {$\phi$};
      \vertex (b) at (-0.5,2) {$\phi$};
      \vertex (c) at (1.5,-2) {$\phi$};
      \vertex (d) at (1.5,2) {$\phi$};
      \vertex (i1) at (1.5,0.5) ;
      \vertex (i2) at (1.5,-0.5) ;
      \vertex (i3) at (2,0) ;
      \vertex (i4) at (1,0) ;
      \vertex (k1) at (-0.5,0.5);
      \vertex (k2) at (-0.5,-0.5);
      \diagram* {
        (a) -- [scalar,momentum=\(p_1\)] (k2) -- [scalar] (k1) -- [scalar, momentum=\(\bar{p}_1\)] (b),
        (c) -- [scalar,momentum=\(p_2\)] (i2) -- [quarter left] (i4) -- [quarter left] (i1) -- [quarter left] (i3) -- [quarter left] (i2),
        (i1) -- [scalar, momentum=\(\bar{p}_2\)] (d),
        (i3) -- [scalar] (i4),
      };
      \fill (i1) circle (2pt);
      \fill (i2) circle (2pt);
      \fill (i3) circle (2pt);
      \fill (i4) circle (2pt);
      \node[circle,inner sep=2pt] (A) at (2.5,2) {};
      \node[circle,inner sep=2pt] (B) at (2.5,-2) {};
    \end{feynman}
  \end{tikzpicture}
  \caption*{Renormalization}
  \end{minipage}
  \begin{minipage}[t]{0.33\textwidth}
    \centering
    \begin{tikzpicture}
    \definecolor{darkgreen}{RGB}{0,100,0}
    \begin{feynman}
      \vertex (a) at (-1,-2) {$\phi$};
      \vertex (b) at (-1,2) {$\phi$};
      \vertex (c) at (1,-2) {$\phi$};
      \vertex (d) at (1,2) {$\phi$};
      \vertex (i1) at (1,0) ;
      \vertex (i2) at (1,-1) ;
      \vertex (i3) at (1.5,-0.5) ;
      \vertex (i4) at (0.5,-0.5) ;
      \vertex (j1) at (1,1.25) ;
      \vertex (j2) at (1,0.25) ;
      \vertex (j3) at (1.5,0.75) ;
      \vertex (j4) at (0.5,0.75) ;
      \vertex (k1) at (-1,0.5);
      \vertex (k2) at (-1,-0.5);

      \diagram* {
        (a) -- [scalar,momentum=\(p_1\)] (k2) -- [scalar] (k1) -- [scalar, momentum=\(\bar{p}_1\)] (b),
        (c) -- [scalar,momentum=\(p_2\)] (i2) -- [quarter left] (i4) -- [quarter left] (i1) -- [quarter left] (i3) -- [quarter left] (i2),
        (i3) -- [scalar] (i4),
        (j2) -- [quarter left] (j4) -- [quarter left] (j1) -- [quarter left] (j3) -- [quarter left] (j2),
        (j1) -- [scalar, momentum=\(\bar{p}_2\)] (d),
      };
      \fill (j1) circle (2pt);
      \fill (i2) circle (2pt);
      \fill (i3) circle (2pt);
      \fill (i4) circle (2pt);
      \node[circle,inner sep=2pt] (A) at (2.5,2) {};
      \node[circle,inner sep=2pt] (B) at (2.5,-2) {};
    \end{feynman}
  \end{tikzpicture}
  \caption*{Four-momentum conservation}
  \end{minipage}
  \caption{Examples of the 4th-order diagrams removed by the redefinition of S-operator, the renormalization technics, and the four-momentum conservation, respectively.}
  \label{fig:removed_diagram}
\end{figure}
\noindent
However, the diagrams in the left and the middle panel in Fig.\ref{fig:removed_diagram} are removed by the redefinition of S-operator and renormalization. 
Furthermore, the diagram in the right panel in Fig.\ref{fig:removed_diagram} is forbidden by the law of four-momentum conservation. Therefore, at the fourth-order, the only diagrams as shown in Fig.\ref{fig:4th-order diagram} are available and
give the amplitude $ {}_{\mathrm{E}}\langle 0 | {}_{\mathrm{s}}\langle \bm{\bar{p}}_1, \bm{\bar{p}}_2 | \ i \hat{T} \ | \bm{p}_1,\bm{p}_2 \rangle_{\mathrm{s}} | 0 \rangle _{\mathrm{E}}$.

\section{Derivation of Eq.\eqref{eq:A_new}}
\label{app:Def_Anew}
In this section, we derive Eq.\eqref{eq:A_new}. To this end, we start with a part of $\mathcal{A}$ in Eq.\eqref{eq:A_def}, 
\begin{align}
    \mathcal{A}_1(p_1,p_2,\bar{p}_1,\bar{p}_2)
    &= \int d^4 q \ D_\mathrm{F}(q) D_\mathrm{F}(q-p_1) D_\mathrm{F}(q-\bar{p}_1) D_\mathrm{F}(q+\bar{p}_2-p_1) \notag \\
    &= \int d^4 q \ \frac{1}{q^2 + m_\text{E}^2 -i\epsilon} \frac{1}{(q-p_1)^2 + m_\text{E}^2 -i\epsilon}\frac{1}{(q-\bar{p}_1)^2 + m_\text{E}^2 -i\epsilon} \frac{1}{(q+\bar{p}_2-p_1)^2 + m_\text{E}^2 -i\epsilon}. 
    \label{eq:def_A_1}
\end{align}
Because of the formula of Feynman parameter integral 
\begin{align}
    \frac{1}{B_1 B_2 \cdots B_n} 
    = (n-1)! \int^{1}_{0} dz_1 \int^{1}_{0} dz_2 &\cdots \int^{1}_{0} dz_n 
    \frac{\delta(1- z_1 - z_2 - \cdots - z_n)}{(z_1 B_1 + z_2 B_2 + \cdots + z_n B_n)^n},
    \label{eq:Int_parameter}
\end{align}
the $\mathcal{A}_1(p_1,p_2,\bar{p}_1,\bar{p}_2)$ is written as follow:
\begin{align}
    \mathcal{A}_1(p_1,p_2,\bar{p}_1,\bar{p}_2)
    &= 3! \cdot \int d^4 q \int^{1}_{0} dz_1 \int^{1}_{0} dz_2 \int^{1}_{0} dz_3 \int^{1}_{0} dz_4 \notag \\
    & \times \frac{ \delta(1 - z_1 - z_2 - z_3 - z_4) }{ \left[ \{(q - p_1)^2 + \tilde{m}^2\}z_1 + (q^2 + \tilde{m}^2)z_2 + \{ (q-\bar{p}_1)^2 + \tilde{m}^2 \}z_3 + \{ ( p_1 - \bar{p}_2 - q )^2 + \tilde{m}^2 \}z_4 \right]^4 },
    \label{eq:Re_A_1}
\end{align}
where $\tilde{m}^2 = m_\text{E}^2 - i\epsilon$. Using the Mandelstam variables $s = (p_1 + p_2)^2$, $t = \ (p_1 - \bar{p}_1)^2$, $u =(p_1 - \bar{p}_2)^2$, we have 
\begin{align}
    &\{(q - p_1)^2 + \tilde{m}^2\}z_1 + (q^2 + \tilde{m}^2)z_2 + \{ (q-\bar{p}_1)^2 + \tilde{m}^2 \}z_3 + \{ ( p_1 - \bar{p}_2 - q )^2 + \tilde{m}^2 \}z_4 \notag \\
    &= \left[ q - \{ z_2 p_1 + z_3 \bar{p}_1 + z_4 ( p_1 - \bar{p}_2 ) \} \right]^2
    + \tilde{m}^2 - \{ z_2 p_1 + z_3 \bar{p}_1 + z_4 (p_1 - \bar{p}_2) \}^2 -z_2 m_\text{s}^2 - z_3 m_\text{s}^2 + z_4 u \notag \\
    &= \tilde{q}^2 + \tilde{m}^2 + (z_2 + z_3)^2 m_\text{s}^2 + z_2 z_3 t + z_1 z_4 u - z_4 u - z_2 m_\text{s}^2 -z_3 m_\text{s}^2 + z_4 u \notag \\
    &= \tilde{q}^2 + \tilde{m}^2 + (z_2 + z_3)(z_2 + z_3 - 1)m_\text{s}^2 + z_2 z_3 t + z_1 z_4 u \notag \\
    &= \tilde{q}^2 + \tilde{m}^2 - (z_1 + z_4)(z_2 + z_3)m_\text{s}^2 + z_2 z_3 t + z_1 z_4 u \notag \\
    &= \tilde{q}^2 + M^2_1-i\epsilon
    \label{eq:Deno_A1}
\end{align}
where $\tilde{q} = q - \{ z_2 p_1 + z_3 \bar{p}_1 + z_4 ( p_1 - \bar{p}_2 ) \} $ and $M^2_1 = m^2_\text{E} - (z_1 + z_4)(z_2 + z_3)m_\text{s}^2 + z_2 z_3 t + z_1 z_4 u$, respectively. Therefore, Eq.\eqref{eq:Re_A_1} is rewritten as
\begin{align}
    \mathcal{A}_1(p_1,p_2,\bar{p}_1,\bar{p}_2)
    &= 3! \cdot \int d^4 \tilde{q} \int^{1}_{0} dz_1 \int^{1}_{0} dz_2 \int^{1}_{0} dz_3 \int^{1}_{0} dz_4 \
    \frac{ \delta(1 - z_1 - z_2 - z_3 - z_4) }{ [ \tilde{q} + M^2_1 -i\epsilon]^4 } \notag \\
    &= 3! \cdot \int^{1}_{0} dz_1 \int^{1}_{0} dz_2 \int^{1}_{0} dz_3 \int^{1}_{0} dz_4 \
    \frac{i\Gamma(2)}{(4\pi)^2 \Gamma(4)} \frac{\delta(1 - z_1 - z_2 - z_3 - z_4)}{[M_1^2 - i\epsilon]^2} \notag \\
    &= \frac{i}{(4\pi)^2} \int^{1}_{0} dz_1 \int^{1}_{0} dz_2 \int^{1}_{0} dz_3 \int^{1}_{0} dz_4 \
    \frac{\delta(1 - z_1 - z_2 - z_3 - z_4)}{[M_1^2 - i\epsilon]^2} 
    \label{eq:Re_A_1-2}
\end{align}
where we performed the $d^4q$ integral. Replacing $M^2_1$ with $M^2_2$ and $M^2_3$, respectively, we find that the other parts in Eq.\eqref{eq:A_def} given by 
\begin{align}
    \mathcal{A}_2(p_1,p_2,\bar{p}_1,\bar{p}_2)
    &= \int d^4 q \ D_\mathrm{F}(q) D_\mathrm{F}(q-p_1) D_\mathrm{F}(q-\bar{p}_1) D_\mathrm{F}(q-p_1-p_2) 
    \label{eq:def_A_2}, 
    \\
    \mathcal{A}_3(p_1,p_2,\bar{p}_1,\bar{p}_2)
    &= \int d^4 q \ D_\mathrm{F}(q) D_\mathrm{F}(q-p_1) D_\mathrm{F}(q+p_2) D_\mathrm{F}(q+\bar{p}_2-p_1) 
    \label{eq:def_A_3}, 
\end{align}
have the following forms
\begin{align}
    \mathcal{A}_2(p_1,p_2,\bar{p}_1,\bar{p}_2)
    &= \frac{i}{(4\pi)^2} \int^{1}_{0} dz_1 \int^{1}_{0} dz_2 \int^{1}_{0} dz_3 \int^{1}_{0} dz_4 \
    \frac{\delta(1 - z_1 - z_2 - z_3 - z_4)}{[M_2^2 - i\epsilon]^2},
    \label{eq:Re_A_2-2}
    \\
    \mathcal{A}_3(p_1,p_2,\bar{p}_1,\bar{p}_2)
    &= \frac{i}{(4\pi)^2} \int^{1}_{0} dz_1 \int^{1}_{0} dz_2 \int^{1}_{0} dz_3 \int^{1}_{0} dz_4 \
    \frac{\delta(1 - z_1 - z_2 - z_3 - z_4)}{[M_3^2 - i\epsilon]^2}.
    \label{eq:Re_A_3-2}
\end{align}
\end{appendix}
Since $\mathcal{A}=\mathcal{A}_1 +\mathcal{A}_2 +\mathcal{A}_3$, using Eqs.\eqref{eq:Re_A_1-2}, \eqref{eq:Re_A_2-2} and \eqref{eq:Re_A_3-2}, we arrive at Eq.\eqref{eq:A_new}.

\end{document}